\documentclass[amsfonts,
	aps,
	prd,
	12pt,
	eqsecnum,
	bibtotoc,
	tightenlines,
	twoside,
	showpacs]{revtex4}

\usepackage{amsmath,epsfig,graphicx,amssymb,amstext}
\usepackage{bm}                         

\newcommand{\wn}{\mathfrak{w}}
\newcommand{\mn}{\mathfrak{m}}
\newcommand{\qn}{\mathfrak{q}}

\def\RE {I\kern-6pt R    }
\def\Z  {Z\kern-13pt Z   }

\def\bi {\begin{itemize} }

\def\ei {\end{itemize}   }
\def\gtwid{\mathrel{\raise.3ex\hbox{$>$\kern-.75em\lower1ex\hbox{$\sim$}}}}
\def\ltwid{\mathrel{\raise.3ex\hbox{$<$\kern-.75em\lower1ex\hbox{$\sim$}}}}
\input epsf

\begin{document}

{\footnotesize \hfill MPP-2007-42}

\title{Isospin diffusion in thermal AdS/CFT with flavor }

\author{Johanna Erdmenger, Matthias Kaminski, Felix Rust}
        \email{jke@mppmu.mpg.de, kaminski@mppmu.mpg.de, rust@mppmu.mpg.de}
\affiliation{Max-Planck-Institut f\"ur Physik (Werner-Heisenberg-Institut),
         F\"ohringer Ring 6,
         80805 M\"unchen, Germany}

\begin{abstract}

We study the gauge/gravity dual of a finite temperature field theory
at finite isospin chemical potential by considering a probe of
two coincident D7-branes embedded in the AdS-Schwarzschild black hole
background. The isospin chemical potential is obtained by giving a vev to the
time component of the non-Abelian gauge field on the brane. 
 The fluctuations of the non-Abelian gauge field on the brane are
dual to the $SU(2)$  flavor current in the field theory.
For the embedding corresponding to vanishing quark mass,
we calculate all Green functions corresponding to the components of
the flavor current correlator. We discuss the physical properties of
these Green functions, which go beyond linear response theory. 
In particular, we show that the isospin chemical 
potential leads to a frequency-dependent isospin diffusion coefficient.  
\end{abstract}


\maketitle
{
        \tableofcontents
}

\section{Introduction}
\label{sec:introduction}
Over the past years,  there have been a number of
lines of investigation for describing QCD-like theories with gravity
duals. In this way, considerable progress towards a gauge/gravity dual
description of phenomenologically 
relevant models has been made. 
One of these lines of investigation
is the gravity dual description of the quark-gluon 
plasma obtained by applying AdS/CFT to relativistic hydrodynamics
\cite{Policastro:2001yc,Policastro:2002se,Son:2002sd,Kovtun:2003wp,Kovtun:2004de,Kovtun:2005ev,Kovtun:2006pf}. The central result of
this approach is the calculation of the shear viscosity from AdS/CFT.  More
recently, an R charge 
chemical potential has been introduced by 
considering gravity backgrounds with R charged black holes
\cite{Son:2006em,Mas:2006dy,Maeda:2006by}, and also the heat conductivity has
been calculated by considering the R current correlators in these backgrounds 
\cite{Son:2006em}. 

A further approach to generalizing the AdS/CFT correspondence to more
realistic field theories  is the addition of flavor to gravity duals via
the addition of probe branes
\cite{Karch:2002sh,Kruczenski:2003be,Babington:2003vm,Kruczenski:2003uq,Sakai:2004cn}. This allows in particular for the calculation of meson masses. 

These two approaches have been combined in order to study the flavor
contribution to finite temperature field theories
from the gravity dual perspective. This began with
\cite{Babington:2003vm} where the embedding of
a D7-brane probe into the AdS-Schwarzschild black hole background was studied
and a novel phase transition was found, which occurs when the D7 probe reaches
the black hole horizon. This transition was shown to be of first order in 
\cite{Kirsch:2004km} (see \cite{Kruczenski:2003uq} for a similar transition in
the D4/D6 system), and studied in further detail in 
\cite{Mateos:2006nu,Albash:2006ew}. Related phase transitions appear
in \cite{Karch:2006bv,Kajantie:2006hv,Schnitzer:2006xz}. 
Mesons in gravity duals of finite temperature field
theories have been studied in
\cite{Ghoroku:2005kg,Ghoroku:2006cc,Peeters:2006iu,Hoyos:2006gb}.

Recently, in view of adding flavor to the quark-gluon plasma, 
the flavor contribution to the shear viscosity has been
calculated in \cite{Mateos:2006yd,Mateos:2007vn}, 
where it was found that $\eta_{\rm fund}
\propto \lambda N_c N_f T^3$.

For a thermodynamical approach in the grand canonical ensemble,
 the inclusion of a
chemical potential and a finite number density is essential. In
\cite{Apreda:2005yz}, an isospin chemical potential was introduced by 
considering two coincident D7 probes, and by giving a vev to the time component
of  the $SU(2)$ gauge field on this probe. This was shown to give rise to a
thermodynamical instability comparable to Bose-Einstein condensation,
compatible with the field-theoretical results of \cite{Harnik:2003ke}. 
For the gauge/gravity dual analysis, a potential
generated by an $SU(2)$ instanton on the D7
probe in the gravity background was used 
\cite{Guralnik:2004ve,Erdmenger:2005bj,Apreda:2005hj,Arean:2007nh}.

A baryon chemical potential $\mu_B$ is obtained by turning on the diagonal
$U(1)\subset U(N_f)$ gauge field on the D7-brane probe \cite{Kobayashi:2006sb}.
Contributions to the D7-brane action arise from the derivative of this
$U(1)$ gauge field with respect to the radial direction.
The effects of this potential on the first-order phase transition described
above have been studied in \cite{Kobayashi:2006sb}, where regions of
thermodynamical instability have been found in the $(T,\mu_B)$ phase
diagram. -- For the $D4/D8/\bar{D8}$ Sakai-Sugimoto model
\cite{Sakai:2004cn,Sakai:2005yt}, the phase transitions in presence of a baryon
number chemical potential, as well as  physical processes such as
photoemission and vector meson screening, 
have been studied in  
\cite{Kim:2006gp,Horigome:2006xu,Parnachev:2006ev}. 

A related approach has been used to calculate the rate of energy loss
of a heavy quark moving through a supersymmetric Yang-Mills plasma 
at large coupling
\cite{Herzog:2006gh}. In this approach the heavy quark is given by a 
classical string attached to the D7-brane probe. -- A first study 
of flavors in thermal AdS/CFT beyond the quenched approximation, i.e. 
with~$N_f\sim N_c$, was performed in~\cite{Bertoldi:2007sf}.

Here we study finite-temperature field theories with
finite isospin chemical potential by considering two coincident
D7-brane probes in the Lorentzian signature AdS-Schwarzschild black hole
background. As in \cite{Apreda:2005yz}, we introduce an isospin chemical
potential by defining
\begin{align} \label{eq:backg} A_0 =
\begin{pmatrix} \mu & 0 \cr 0 & -\mu \end{pmatrix} \, , 
\end{align}
for the time component of the $SU(2)$ gauge field on the two coincident 
D7-branes.  This constant chemical potential is a solution to the D7-brane
equations of motion and is present even for the D7-brane embedding 
corresponding to massless quarks. We consider small $\mu$, such that
the Bose-Einstein instability mentioned above, which is 
of order ${\cal{O}} (\mu^2)$,
does not affect our discussion here. 

For simplicity we consider only the D7 probe embedding for which the
quark mass vanishes, $m=0$. This embedding is constant and terminates
at the horizon. We establish the $SU(2)$ non-Abelian action for a
probe of two coincident D7-branes. We obtain 
the equations of motion for fluctuations about the background given by
(\ref{eq:backg}). These are dual to the $SU(2)$ flavor current
$J^{\mu a}$.  We find an ansatz for decoupling the equations of motion
for the different Lorentz and flavor components, and solve them by
adapting the method developed in 
\cite{Policastro:2002se,Son:2002sd}. This involves Fourier transforming to
momentum space, and using a power expansion ansatz for the equations of motion.
We discuss the approximation necessary for an analytical
solution, which amounts to considering frequencies with
$\omega<\mu<T$.  With this approach we obtain the complete current-current
correlator. The key point is that the constant chemical potential
effectively replaces a time derivative in the action and in the
equations of motion. In the Fourier transformed picture, this leads to
a square-root dependence of physical observables on the frequency,
$\sqrt{\omega}$. This non-linear behavior goes beyond
linear response theory. We discuss the physical properties of the Green
functions contributing to the current-current correlator.  In
particular, for small frequencies 
we find a frequency-dependent diffusion coefficient
$D(\omega) \propto \frac{1}{T} \sqrt{\omega/ \mu}$. Whereas
frequency-dependent diffusion has -- to our knowledge --
 not yet been discussed in the
context of the quark-gluon plasma, it is well-known in the theory of
quantum liquids. For instance, for small frequencies the square-root 
behavior we find agrees qualitatively
with the results of \cite{PhysRevLett,Rabani} for
liquid para-hydrogen. Generally, frequency-dependent diffusion leads to a
non-exponential decay of time-dependent fluctuations, as discussed for a
classical fluid in \cite{Bhattacharjee:1980tf}.

Physically, the isospin chemical potential corresponds to the energy
necessary to inverting the isospin of a given particle. Within nuclear
physics, such a chemical potential is of relevance for the description
neutron stars. Moreover,
isospin diffusion has 
been measured in heavy ion reactions~\cite{Liu:2006xs,Tsang:2003td}.
-- For two-flavor QCD, effects of a finite isospin chemical potential
have been discussed for instance in
\cite{Son:2000xc,Splittorff:2000mm,Toublan:2003tt}. 
The phase diagrams discussed there are beyond the
scope of the present paper. We expect to return to similar diagrams in
the gauge/gravity dual context in the future. 

This paper is organized as follows. In section 2 we summarize the
AdS/CFT hydrodynamics approach to calculating Green functions, 
which we use in the subsequent. Moreover we comment on
frequency-dependent diffusion within hydrodynamics. 
In section 3 we establish the D7 probe
action in presence of the isospin chemical potential, derive the
corresponding equations of motion and solve them. In section 4 we
obtain the associated Green functions in the hydrodynamical
approximation. We discuss their pole structure and obtain the
frequency-dependent diffusion coefficient. We conclude 
in section 5 with an interpretation of our results. An explanation of our notation
as well as a series of calculations are relegated to a number of appendices.

\section{Hydrodynamics and AdS/CFT}
\label{sec:hydroAdS}
Thermal Green functions have proven to be a useful tool 
for analyzing the structure of hydrodynamic 
theories and for calculating hydrodynamic quantities such as
transport coefficients. For instance, given a 
retarded current correlation function~$G(\vec k)_{\mu \nu}$ in Minkowski
space,  the spectral function can
be written in terms of its imaginary part,
\begin{equation}        
\label{eq:spectralFunction}
\chi_{\mu\nu}(\vec k)=\, -2\, \mathrm{Im} G_{\mu\nu}(\vec k) \, .
\end{equation}  
For the gravity dual approach, 
this is discussed for instance in \cite{Kovtun:2006pf,Teaney:2006nc}. 

In this paper we use the gauge/gravity dual 
prescription of \cite{Son:2002sd} 
for calculating Green functions 
in Minkowski spacetime.  For further reference, we outline this
prescription in the subsequent. It is based on the AdS/CFT-correspondence
relating supergravity fields~$A$ in a black hole background
to operators~$J$ in the dual
gauge theory. The black hole background is asymptotically Anti-de
Sitter space and places the dual field theory at finite
temperature. This temperature
corresponds to the Hawking temperature of the black hole, or more
generally speaking, of the black branes. 
Starting out from a classical supergravity
action~$S_{\rm cl}$ for the gauge field $A$, according to \cite{Son:2002sd}
we extract the function~$B(u)$~(containing metric factors and
the metric determinant)
in front of the kinetic term~$(\partial_u A)^2$,
\begin{equation}
\label{eq:classicalAction}
S_{\mathrm{cl}}= \,\int \mathrm d u \mathrm d^4x\,
  B(u)\,(\partial_u A)^2\, +\, \dots
\end{equation}
Then we perform a Fourier transformation and solve the
linearized equations of motion in momentum space. This is a
second order differential equation, so we have to fix two boundary
conditions. The first one at the boundary of AdS at~$u=0$ can
be written as
\begin{equation}
\label{eq:uIs0Boundary}
A(u,\vec k)\,=\, f(u,\vec k)\, A^{\text{bdy}}(\vec k) \, ,
\end{equation}
where~$A^{\text{bdy}}(\vec k)$ is the value of the supergravity field at
the boundary of AdS depending only on the four flat
boundary coordinates. Thus by definition we
have~$\left. f(u,\vec k)\right|_{u\to0}=1$.
For the other boundary, located at the horizon~$u=1$, we impose
the incoming wave condition. This requires that
any Fourier mode~$A(\vec k)$ with timelike~$\vec k$
can travel into the black hole, but is not allowed to cross
the horizon in the opposite direction. For spacelike $\vec k$, the components
of $A$ have to be regular at the horizon.
Then the retarded thermal Green function is given by
\begin{equation}
\label{eq:retardedThermalGreen}
G(\omega, {\bm q})\,=\, \left. -2\, B(u)\, f(u,-\vec k)\,
  \partial_u\,f(u,\vec k)\right|_{u\to 0} \, .
\end{equation}

The thermal correlators obtained in this way display
hydrodynamic properties, such as
poles located at complex frequencies. Generically, for the R current component 
correlation functions calculated from supergravity, there are retarded
contributions of the form
\begin{equation}
G(\omega, {\bm q}) \propto \frac{1}{i \omega - D {\bm q}^2} \, .
\end{equation}
This may be identified with the 
the Green function for the hydrodynamic diffusion equation
\begin{equation}
\label{eq:diffusionEquation}
\partial_0\, J_0 (t, \bm{x}) \,=\, D \, \nabla^2\, J_0(t, \bm{x}) \, ,
\end{equation}
with $J_0$ the time component of a diffusive current.  $D$ is the diffusion
constant. In Fourier space this equation reads
\begin{equation} \label{eq:constantD}
i \omega J_0(\omega, \bm{ k}) = D \bm{ k}^2 J_0(\omega, {\bm k} ) \, .
\end{equation}
In position space, this corresponds to an exponential decay of $J_0$ with time.

For the non-Abelian case with an isospin chemical potential, in
sections \ref{sec:sugraBackAndAction} and \ref{sec:heatAndDiff} we will obtain
retarded Green functions of the form
\begin{equation} \label{eq:omegadiff}
G(\omega, {\bm q}) \propto \frac{1}{i \omega - D(\omega)  {\bm q}^2} \, .
\end{equation}
Retarded Green functions of this type have been discussed for instance in
\cite{Bhattacharjee:1980tf}. 
(\ref{eq:omegadiff})  corresponds
 to frequency-dependent diffusion
with coefficient $D(\omega)$, such that (\ref{eq:constantD}) becomes
\begin{equation} \label{eq:omegaD}
i \omega J_0(\omega, {\bm k}) = D (\omega)  \bm{k}^2 J_0(\omega, \bm{k}) \, .
\end{equation}
In our case, $J_0$ is 
the averaged isospin at a given point in the liquid. 

This is a non-linear behavior which goes beyond linear response theory.
In particular, when
Fourier-transforming back to position space, we have to use the
convolution for the product $D \cdot J_0$ and obtain
\begin{equation} \label{eq:convolution}
\partial_0 J_0 (t,\bm{x}) + \nabla^2 \int\limits_{-\infty}^{t} \mathrm{d}s\,
J_0(s,\bm{x}) D(t-s) = 0
\end{equation}
for the redarded Green function.
This implies together with the continuity equation 
$\partial_0\, J_0\,+ \bm{\nabla}\cdot \bm{J}\,=\,0 $, with $\bm{J}$
the three-vector current associated to $J_0$, that 
\begin{equation}
{\bm J} = - {\bm \nabla} (D*J_0) \, ,
\end{equation}
where $*$ denotes the convolution. This replaces the linear response
theory constitutive equation ${\bm J} = - D {\bm{\nabla}}  J_0 $. 
Note that for $D(t-s)= D \delta(t-s)$ with $D$ constant, 
(\ref{eq:convolution})  reduces again to
(\ref{eq:diffusionEquation}).

\section{Supergravity background and action}
\label{sec:sugraBackAndAction}

\subsection{Finite temperature background and brane configuration}
\label{sec:backAndBranes}

We consider an asymptotically $AdS_5\times S^5$ spacetime as the near horizon limit
of a stack of $N_c$ coincident D3-branes. More precisely, as in
\cite{Policastro:2002se}, our background is an $AdS$ black hole, which is the
geometry dual to a field theory at finite temperature. The Minkowski
signature background is
\begin{equation}
\label{eq:AdSBHmetric}
\begin{gathered}
\mathrm{d}s^2= \frac{b^2 R^2}{u} \left(-f(u)\,\mathrm d x_0^2+
  \mathrm d x_1^2+\mathrm d x_2^2+\mathrm d x_3^2 \right) +  
  \frac{R^2}{4 u^2 f(u)}\, \mathrm d u^2 + R^2 \mathrm d \Omega_5^2,\\
  0 \leq u \leq 1,\qquad x_i\in\mathbb{R}, \qquad C_{0123}=\frac{b^4 R^4}{u^2},
\end{gathered}
\end{equation}
with the metric $\mathrm d \Omega_5^2$ of the unit $5$-sphere, and the function
$f(u)$, $AdS$ radius $R$ and temperature parameter $b$ given in terms of the
string coupling $g_s$, temperature $T$, inverse string tension $\alpha'$
and number of colors $N_c$ by
\begin{equation}
\label{eq:metricDefinitions}
f(u)=1-u^2,\qquad R^4=4\pi g_s N_c  {\alpha'}^2,\qquad b=\pi T.
\end{equation}
The geometry is asymptotically $AdS_5\times S^5$ with the boundary of the $AdS$
part located at $u=0$. At the black hole horizon the radial coordinate
$u$ has the value $u=1$.

Into this ten-dimensional spacetime we embed $N_f=2$ coinciding D7-branes,
hosting flavor gauge fields $A_\mu$.  The embedding we choose
extends the D7-branes in all directions of $AdS$ space and wraps an $S^3$ on the
$S^5$. In this work we restrict ourselves to the most straightforward case, that
is the embedding of the branes through the origin along the $AdS$ radial
coordinate $u$. This corresponds to massless quarks in the dual field theory. On
the brane, the metric in this case simply reduces to
\begin{equation}
\label{eq:finiteTMetric} 
\begin{gathered}
\mathrm{d}s^2= \frac{b^2 R^2}{u} \left(-f(u)\,\mathrm d x_0^2+
  \mathrm d x_1^2+\mathrm d x_2^2+\mathrm d x_3^2 \right) +  
  \frac{R^2}{4 u^2 f(u)}\, \mathrm d u^2 + R^2 \mathrm d \Omega_3^2,\\
  0 \leq u \leq 1,\qquad x_i\in\mathbb{R}.
\end{gathered}
\end{equation}
Due the choice of our gauge field in the next subsection, the
remaining three-sphere in this metric will not play a prominent role.

The table below gives an overview of the indices we use to refer to 
certain directions and subspaces.

\begin{center}
	\vspace{2.5ex}
	\includegraphics{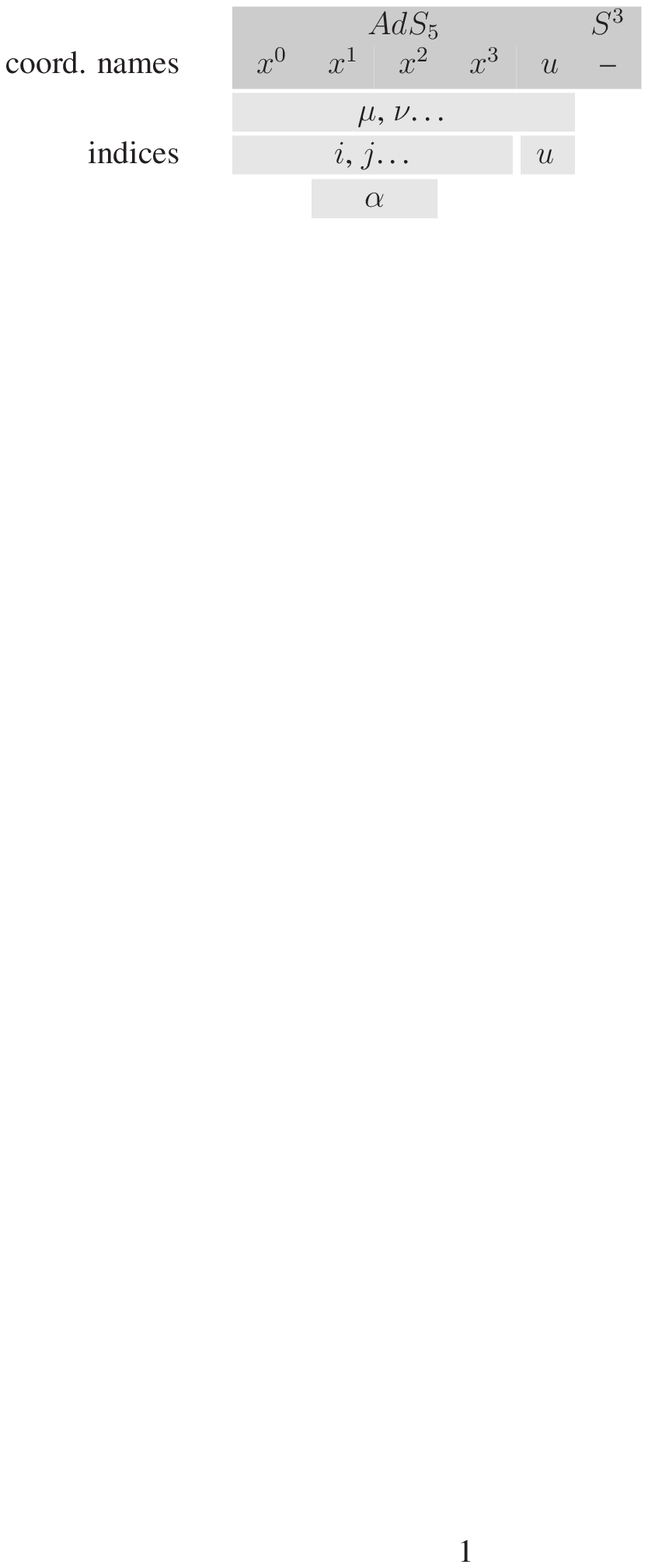}
\end{center}

\subsection{Introducing a non-Abelian chemical potential}

A gravity dual description of a
chemical potential amounts to a non-dynamical time component of the
gauge field $A_\mu$ in the action for the D7-brane probe embedded into
the background given above.
There are essentially two 
different ways to realize a non-vanishing contribution from a chemical
potential to the
field strength tensor $F=2\partial_{[\mu}A_{\nu]}+f^{abc}A^b_\mu A^c_\nu$. 
The first is to consider a $u$-dependent baryon chemical potential for
a single brane probe. The second, which we pursue here, is to consider
a constant isospin chemical potential. This requires a non-Abelian
probe brane action and thus a probe of at least two coincident
D7-branes,  as suggested in
\cite{Apreda:2005yz}. Here the time component of the gauge field is
taken to be
\begin{equation}
\label{eq:nonAbelianMu}
A_0^{\vphantom{a}} = A^a_0 \, t^a ,
\end{equation}
where we sum over indices which occur twice in a term and denote the
gauge group generators by $t^a$. The brane configuration described
above corresponds to an $SU(N_f)$ gauge group with $N_f=2$ on the
brane, which corresponds to a global $SU(N_f)$ in the dual field theory. 
For $N_f=2$, the generators of the gauge group on the brane 
are given by $t^a=\frac{\sigma^a}{2}$, with Pauli matrices $\sigma^a$.
We will see that (\ref{eq:nonAbelianMu}) indeed produces non-trivial new
contributions to the
action.

Using the standard background field method of quantum field theory, we will
consider the chemical potential as a fixed background and let the gauge fields
fluctuate. We single out a particular direction in flavor space by taking
$A^3_0=\mu$ as the only non-vanishing component of the background field. From
now on we use the symbol $A^a_\nu$ to refer to gauge field fluctuations around
the fixed background,
\begin{equation}
\label{eq:backgroundRule}
A^a_\nu \to \mu \delta_{\nu 0}\delta^{a3} + A^a_\nu \, .
\end{equation}

We pick the gauge in which $A_u\equiv 0$ and  assume that
$A_\mu \equiv 0$ for $\mu=5,6,7$. Due to the symmetries of the
background,
we effectively examine gauge field
fluctuations $A_\mu$ living in the five-dimensional subspace on
the brane spanned by the coordinates $x_0$, $x_1$, $x_2$, $x_3$ and by the
radial AdS coordinate $u$. 
The magnitude of all components of $A$ and the
background chemical potential 
$\mu$ are considered to be small. This allows us to simplify
certain expressions by dropping terms of higher order in $A$ and in the
chemical potential $\mu$.

\subsection{Dirac-Born-Infeld action}

The action describing the dynamics of the flavor gauge fields in
the setup of this work is the Dirac-Born-Infeld action. There are no 
contributions from the Chern-Simons action, which would require
non-zero gauge field components in all of the 4,5,6,7-directions. 
As mentioned, we consider 
the D7 probe embedding whose asymptotic value at the boundary is
chosen such that it corresponds to vanishing quark mass, $m=0$. The
metric on the brane is then given by (\ref{eq:finiteTMetric}).
Since we are interested in two-point correlators only, it is sufficient to
consider the action to second order in $\alpha'$,
\begin{equation}
\label{eq:dbiAction}
S_{\text{D7}} = -T_{7} \frac{(2 \pi \alpha')^2}{2} 2\pi^2 R^3 
  \, T_r \int\limits_{u_b=0}^{u_h=1} \mathrm d u\,
  \mathrm d^4 x\: \sqrt{-g}\,g^{\mu\mu'}\,g^{\nu\nu'}\, F_{\mu\nu}^a
  \, F_{\mu'\nu'}^a \, ,
\end{equation}
where we use the following definitions for the D7-brane tension $T_7$ and the
trace over the representation matrices $t^a$,
\begin{align}
\label{eq:actionDefinitions}
T_7 &= (2\pi)^7 g_s^{-1} (\alpha')^{-4} \, , \\
\mathrm{tr} (t^a\, t^{b}) &= T_r\, \delta^{ab} \, .
\end{align}
In our case we have $T_r=1/2$. The overall factor $2\pi^2 R^3$ comes from the
integration over the $5,6,7$-directions, which are the directions along
the $S^3$. 

Evaluating the DBI action given in~(\ref{eq:dbiAction}) with
the substitution rule~(\ref{eq:backgroundRule}), we arrive at
\begin{equation}
\begin{aligned}
\label{eq:dbiActionWithMu}
S_{{\mathrm D7}} =\;& -T_{7} \frac{(2 \pi \alpha')^2}{2} 2\pi^2 R^3 \, T_r\\
        & \times\int\limits_{u_b=0}^{u_h=1} \mathrm d u\,
  \mathrm d^4 x\: \sqrt{-g}g^{\mu\mu'}g^{\nu\nu'}\left(
  4 \partial_{[\mu}^{\vphantom{a}} A^a_{\nu]}\,
  \partial_{[\mu'}^{\vphantom{a}} A^a_{\nu']}  -
  8 \delta_{0\nu} \delta_{0\nu'} f^{abc} \partial_{[0}^{\vphantom{a}} A^a_{\mu]}\,
  A^b_{\mu'}\, \mu^c  \right),
\end{aligned}
\end{equation}
where we use the short-hand notation $\mu^c=\mu \delta^{3c}$ and
neglect terms of higher than linear order in $\mu$, and higher than quadratic
order in $A$ since both are small in our approach.

Up to the sum over flavor indices $a$,
the first term in the bracket in (\ref{eq:dbiActionWithMu}) 
is reminiscent of the Abelian super-Maxwell action in five dimensions,
considered already for the R charge current correlators 
in~\cite{Policastro:2002se}.
The new second term in
our action arises from the non-Abelian nature of the gauge group, giving
terms proportional to the gauge group's structure constants $f^{abc}$ in the
field strength tensor 
$F^a_{\mu\nu}=2\partial_{[\mu}^{\vphantom{a}}A^a_{\nu]}+f^{abc} A^b_\mu\, A^c_{\nu}$.

\subsection{Equations of motion}
\label{sec:eom}

We proceed by calculating the retarded Green functions for the action
(\ref{eq:dbiActionWithMu}), following the prescription of 
\cite{Son:2002sd} as outlined in section 2 above. According to this
prescription, as a first step we consider the equations of motion 
obtained from the
action~(\ref{eq:dbiActionWithMu}), which are given by
\begin{equation}
\begin{aligned}
\label{eq:eom}
0 =\; & 2 \partial_{\mu}\left(
  \sqrt{-g}\, g^{\mu\mu'} g^{\nu\nu'}\: \partial^{\vphantom{a}}_{[\mu'} A^a_{\nu']}
  \right)  \\
  &+ f^{abc}\left[ \sqrt{-g} g^{00} g^{\nu\nu'}\:\mu^c
   \left(\partial^{\vphantom{b}}_{\nu'} A_{0}^b-2\partial^{\vphantom{b}}_0 A_{\nu'}^b\right)
  +\delta^{0\,\nu} \partial^{\vphantom{b}}_\mu\left(
  \sqrt{-g}\, g^{00} g^{\mu\mu'} A_{\mu'}^b \mu^c \right)\right].
\end{aligned}
\end{equation}
It is useful to work in momentum space from now on. 
We therefore expand the bulk
gauge fields in Fourier modes in the $x^i$ directions,
\begin{equation}
\label{eq:fourierTrafo}
A_\mu (u,\vec{x})=\int \frac{\mathrm d^4 k}{(2\pi)^4}
  e^{-i \omega x_0 + i \bm{k}\cdot \bm{x}} A_\mu(u,\vec{k}).
\end{equation}
As we work in the gauge where $A_u=0$, we only have to take care of the
components $A_i$ with $i=0,1,2,3$.

For the sake of simplicity, we choose the momentum of the fluctuations to be
along the $x_3$ direction, so their momentum four-vector is 
$\vec{k}=(\omega,0,0,q)$.
With this choice we have specified to gauge fields which
only depend on the radial coordinate $u$, the time coordinate
$x_0$ and the spatial $x_3$ direction.

\subsubsection{Equations for $A_1^a$- and $A_2^a$-components}

Choosing the free Lorentz index in the equations of motion~(\ref{eq:eom})
to be $\nu=\alpha=1,2$ gives two identical differential equations for $A_1$ and
$A_2$,
\begin{equation}
\label{eq:eomA1A2}
0= {A^a_\alpha}''+\frac{f'}{f}{A^a_\alpha}'+
   \frac{\wn^2-f \qn^2 }{u f^2}A^a_\alpha+
   2i\frac{\wn}{u f^2} f^{abc} \frac{\mu^b}{2\pi T} A^c_\alpha,
\end{equation}
where we indicated the derivative with respect to $u$ with a prime and have
introduced the dimensionless quantities
\begin{equation}
\label{eq:definitionWQM}
\wn=\frac{\omega}{2 \pi T} \, ,
\qquad \qn=\frac{q}{2 \pi T} \, ,\qquad \mn=\frac{\mu}{2 \pi T} \, .
\end{equation}
We now make use of the structure constants of $SU(2)$, which are 
$f^{abc}=\varepsilon^{abc}$, where $\varepsilon^{abc}$ is the totally antisymmetric
epsilon symbol with $\varepsilon^{123}=1$. Writing out (\ref{eq:eomA1A2}) for
the three different choices of $a=1,2,3$ results in
\begin{eqnarray}
\label{eq:eomA1A2Flavor1}
0&=& {A_\alpha^1}''+\frac{f'}{f} {A_\alpha^1}'
  +\frac{\wn^2 - f \qn^2}{u f^2} A_\alpha^1-
  2i\frac{\mn\wn}{u f^2} A_\alpha^2 \, ,    \\
\label{eq:eomA1A2Flavor2}
0&=& {A_\alpha^2}''+\frac{f'}{f} {A_\alpha^2}'
  +\frac{\wn^2 - f \qn^2}{u f^2} A_\alpha^2+
  2i\frac{\mn\wn}{u f^2} A_\alpha^1 \, ,    \\
\label{eq:eomA1A2Flavor3}
0&=& {A_\alpha^3}''+\frac{f'}{f} {A_\alpha^3}'
  +\frac{\wn^2 - f \qn^2}{u f^2}  A_\alpha^3 \, .
\end{eqnarray}
The first two of these equations are coupled, the third one
is the same equation that was solved in the Abelian
Super-Maxwell case~\cite{Policastro:2002se}.

\subsubsection{Equations for $A_0^a$- and $A_3^a$-components}

The remaining choices for the free Lorentz
index $\nu=0,3,u$ in~(\ref{eq:eom}) result in
three equations which are not independent. The choices $\nu=0$ and $\nu=u$ give
\begin{eqnarray}
\label{eq:eomA0A3nuIs0}
0&=& {A_0^a}''-\frac{\qn^2}{u f}A_0^a-\frac{\qn\wn}{u f}A_3^a
  - i \frac{\qn}{u f} f^{abc} \frac{\mu^b}{2 \pi T}A_3^c \, ,  \\
\label{eq:eomA0A3nuIsu}
0&=& \wn {A_0^a}'+\qn f {A_3^a}' + i f^{abc} \frac{\mu^b}{2\pi T}
{A_0^c}' \, .
\end{eqnarray}
Solving~(\ref{eq:eomA0A3nuIsu}) for ${A_0^a}'$, differentiating
it once with respect to $u$ and using (\ref{eq:eomA0A3nuIs0}) results in
equation (\ref{eq:eom}) for $\nu=3$,
\begin{equation}
\label{eq:eomA0A3nuIs3}
0= {A_3^a}''+\frac{f'}{f}{A_3^a}'+\frac{\wn^2}{u f^2}\, A_3^a+
  \frac{\qn\wn}{u f^2}A_0^a
  +  i \frac{\qn}{u f^2} f^{abc} \frac{\mu^b}{2\pi T}\, A_0^c
  + 2i \frac{\wn}{u f^2} f^{abc} \frac{\mu^b}{2\pi T}\, A_3^c \, .
\end{equation}
We will make use of the equations (\ref{eq:eomA0A3nuIs0}) and
(\ref{eq:eomA0A3nuIsu}) which look more concise.  These equations of motion for
$A_0^a$ and $A_3^a$ are coupled in Lorentz and flavor indices. To decouple them
with respect to the Lorentz structure, we solve (\ref{eq:eomA0A3nuIsu}) for
${A_3^a}'$ and insert the result into the differentiated version of 
(\ref{eq:eomA0A3nuIs0}). This gives
\begin{equation}
\label{eq:eomA0}
        0 = {A^a_0}''' + \frac{(uf)'}{uf}\,{A^a_0}'' + \frac{\wn^2-f\qn^2}{uf^2}\,{A^a_0}'
             + 2i\frac{\wn}{uf^2}\,f^{abc} \frac{\mu^b}{2 \pi T}\, {A^c_0}'.
\end{equation}

The equations for $a=1,2$ are still coupled with respect to their gauge 
structure. The case $a=3$ was solved in \cite{Policastro:2002se}.
We will solve (\ref{eq:eomA0}) for ${A^a_0}'$ and can obtain ${A^a_3}'$ from
(\ref{eq:eomA0A3nuIsu}). 
Note that it is sufficient for our purpose to obtain solutions for
the derivatives of the fields. These contribute to 
(\ref{eq:retardedThermalGreen}), while the functions
$A=f(u,\vec{k}) A^{\text{bdy}}(\vec{k})$ themselves simply
contribute a factor of $f(u,-\vec{k})$ which is one at the boundary.

\subsubsection{Solutions}
\label{sec:solutions}

Generally, we follow the methods developed in~\cite{Policastro:2002se},
since our differential equations are very similar to the ones considered there.
Additionally, we need to respect the flavor structure of the gauge fields.  The
equations for flavor index $a=3$ resemble the ones analyzed in
\cite{Policastro:2002se}.  However, those for $a=1,2$ involve extra
terms, which
couple these equations. In our case the equations are coupled not
only via their Lorentz indices, but also with respect to the flavor indices. 
We already decoupled the Lorentz structure in the previous section.
As shown below, the equations of motion which involve different gauge
components will decouple if we consider the variables
\begin{equation}
\label{eq:flavorTrafo}
\begin{aligned}
X_i            & =&& A_i^1 + i A_i^2,\\
\widetilde X_i & =&& A_i^1 - i A_i^2.
\end{aligned}
\end{equation}
Here the $A_i^1$, $A_i^2$ 
are the generally complex gauge field components in momentum
space. Note that up to $SU(2)$ transformations, the combinations
(\ref{eq:flavorTrafo}) are the only ones which decouple the equations
of motion for $a=1,2$.  
These combinations are reminiscent of the non-Abelian $SU(2)$ gauge field
in position space,
\begin{equation}
\label{eq:gaugeFieldMatrix}
A_i= A^a_i \frac{\sigma^a}{2}
= \frac{1}{2}
\begin{pmatrix}
A^3_i & A^1_i - i A^2_i \\
A^1_i + i A^2_i  & - A^3_i \\
\end{pmatrix}.
\end{equation}
The equations of motion for the flavor index $a=3$ were solved in
\cite{Policastro:2002se}. To solve the equations of motion for the fields
$A^a_i$ with $a=1,2$, we rewrite them in terms of $X_i$ and
$\widetilde X_i$.
Applying the transformation~(\ref{eq:flavorTrafo}) to the equations of
motion (\ref{eq:eomA1A2Flavor1}) and (\ref{eq:eomA1A2Flavor2}) and the
$a=1,2$ versions of (\ref{eq:eomA0}) and (\ref{eq:eomA0A3nuIsu}) leads to
\begin{eqnarray}
\label{eq:eomXalpha}
0 & = & X_\alpha'' +\frac{f'}{f} X_\alpha'
        +\frac{\wn^2-f \qn^2 \mp 2 \mn\wn}{u f^2} X_\alpha,\quad\qquad \alpha=1,2,\\
\label{eq:eomX0}
0 & = & X_0''' + \frac{(uf)'}{uf}\,X_0'' 
            + \frac{\wn^2-f\qn^2\mp 2\mn\wn}{uf^2} X_0',\\
\label{eq:eomX3}
0 & = & \left(\wn \mp \mn\right) X_0' + \qn f X_3',
\end{eqnarray}
where the upper signs correspond to $X$ and the lower ones to $\widetilde X$.

We see that some coefficients of these functions are divergent
at the horizon $u=1$. Such differential equations with
singular coefficients are generically solved by an ansatz
\begin{equation}
\label{eq:ansatzA1A2}
        X_i=(1-u)^\beta\, F(u), \qquad 
        \widetilde X_i=(1-u)^{\widetilde \beta}\, \widetilde F(u),
\end{equation}
with regular functions $F(u)$ and $\widetilde F(u)$. To cancel the singular
behaviour of the coefficients, we have to find the adequate $\beta$ and 
$\widetilde \beta$, the so-called indices, given by equations known as the
indicial equations for
$\beta$ and $\widetilde \beta$. We eventually get for all $X_i$ and
$\widetilde X_i$
\begin{equation}
\label{eq:indices}
        \beta=\pm\frac{i\wn}{2}\sqrt{1-\frac{2\mn}{\wn}}, \qquad
        \widetilde \beta=\pm\frac{i\wn}{2}\sqrt{1+\frac{2\mn}{\wn}}.
\end{equation}

Note that these exponents differ from those of the
Abelian Super-maxwell 
theory~\cite{Policastro:2002se} by a dependence on
$\sqrt{\wn}$ in the limit of
small frequencies~($\wn<\mn$). In the limit of vanishing chemical
potential~$\mn\to 0$, the indices given in~\cite{Policastro:2002se}
are reproduced from~(\ref{eq:indices}).
In order to solve (\ref{eq:eomXalpha}),~(\ref{eq:eomX0})
and~(\ref{eq:eomX3}), we wish to
introduce a series expansion ansatz in the momentum variables~$\wn$
and~$\qn$.
In fact, the physical motivation behind this expansion is that
we aim for thermodynamical quantities which are known from
statistical mechanics in the hydrodynamic limit of
small four-momentum $\vec{k}$.
So the standard choice would be
\begin{equation}
\label{eq:standardHydroLimit}
F(u)= F_0 + \wn F_1 + \qn^2 G_1 + \ldots \, .
\end{equation}
On the other hand, we realize that our indices will appear
linearly (and quadratically) in the differential equations'
coefficients after inserting~(\ref{eq:ansatzA1A2})
into~(\ref{eq:eomXalpha}),~(\ref{eq:eomX0})
and~(\ref{eq:eomX3}). The square root in~$\beta$
and~$\widetilde \beta$ mixes different orders of~$\wn$.
In order to sort coefficients in our series ansatz, we assume $\wn<\mn$ and
keep only the leading $\wn$ contributions to $\beta$ and $\widetilde \beta$,
such that
\begin{equation}
\label{eq:approximateIndices}
\beta\approx \mp \sqrt{\frac{\wn\mn}{2}},\qquad
\widetilde \beta\approx \pm i \sqrt{\frac{\wn\mn}{2}}.
\end{equation}
This introduces an additional order~$\mathcal{O}(\wn^{1/2})$,
which we include in our ansatz~(\ref{eq:standardHydroLimit})
yielding
\begin{equation}
\label{eq:hydroLimit}
F(u)= F_0 + \wn^{1/2} F_{1/2} + \wn F_1 + \qn^2 G_1 + \ldots\, ,
\end{equation}
and analogously for the tilded quantities. If we had not included
$\mathcal{O}(\wn^{1/2})$ the resulting system would be overdetermined.
The results we obtain by using the approximations (\ref{eq:approximateIndices})
and (\ref{eq:hydroLimit}) have been checked against the
numerical solution for exact $\beta$ with exact~$F(u)$. These approximations
are useful for fluctuations
with $\qn,\wn < 1$ (see subsection~\ref{sec:compareAnaNum} in
Appendix~\ref{sec:solutionsEOM}). Note that by dropping the $1$
in~(\ref{eq:indices}) we also drop the Abelian limit.

Consider the indices~(\ref{eq:indices}) for positive frequency first.
In order to meet the incoming wave boundary condition introduced in
section~\ref{sec:hydroAdS}, we restrict the solution $\widetilde \beta$
to the negative sign only. For the approximate~$\widetilde \beta$
in~(\ref{eq:approximateIndices}) we therefore choose the lower (negative) sign.
This exponent describes a mode that travels into the horizon of the
black hole. In case of $\beta$ we demand the mode to
decay towards the horizon, choosing the lower (positive) sign
in~(\ref{eq:approximateIndices}) consistently.
Note that for negative frequencies~$\omega <0$ the indices $\beta$
and~$\widetilde \beta$ exchange their roles.

Using (\ref{eq:approximateIndices}) in (\ref{eq:ansatzA1A2}) and
inserting the ansatz into the equations of motion, we find equations 
for each order in $\qn^2$ and $\wn$ separately. 
After solving the equations of motion for the coefficient functions $F_0$,
$F_{1/2}$, $F_1$ and $G_1$, we eventually can assemble
the solutions to the equations of motion for $X$ as defined in
(\ref{eq:flavorTrafo}), 
\begin{equation}
\label{eq:fullX}
        X(u) = (1-u)^\beta \, F(u) = (1-u)^\beta \, 
  \left( F_0 + \sqrt{\wn} F_{1/2} + \wn F_1 + \qn^2 G_1 + \ldots \right).
\end{equation}
and a corresponding formula for $\widetilde X(u)$ from the ansatz
(\ref{eq:ansatzA1A2}).

\bigskip

Illustrating the method, we now write down the equations of motion order by
order for the function $X_\alpha$. To do so, we use
 (\ref{eq:fullX}) with (\ref{eq:approximateIndices}) in
(\ref{eq:eomXalpha}) with the upper sign for $X_\alpha$. Then we 
examine the result
order by order in $\wn$ and $\qn^2$,
\begin{align}
\label{eq:eomFSortedByOrders1}
\mathcal {O}(\text{const}):\qquad
0&= F_0''+\frac{f'}{f} F_0' \, ,\\
\mathcal{O}(\sqrt{\wn}):\qquad
\label{eq:eomFSortedByOrders2}
0&= F_{1/2}''+\frac{f'}{f}F_{1/2}' -\frac{\sqrt{2\mn}}{1-u} F_0'
  -\sqrt{\frac{\mn}{2}}\frac{1}{f} F_0\, ,\\
\mathcal{O}(\wn):\qquad
\label{eq:eomFSortedByOrders3}
0&= F_1''+\frac{f'}{f} F_1'-\frac{\sqrt{2\mn}}{1-u}F_{1/2}'
  - \sqrt{\frac{\mn}{2}} \frac{1}{f} F_{1/2}
  - \mn\frac{4-u(1+u)^2}{2 u f^2} F_0 \, ,\\
\label{eq:eomFSortedByOrders4}
\mathcal{O}(\qn^2):\qquad
0&= G_1''+\frac{f'}{f} G_1'- \frac{1}{u f} F_0 \, .
\end{align}

At this point we observe that the differential equations we have to solve for
each order are shifted with respect to the solutions found in
\cite{Policastro:2002se}. The contributions of order $\wn^n$ in
\cite{Policastro:2002se} now show up in order $\wn^{n/2}$. Their
solutions will exhibit factors of order $\mu^{n/2}$.

Solving the system (\ref{eq:eomFSortedByOrders1}) to
(\ref{eq:eomFSortedByOrders4}) of coupled differential equations is
straightforward in the
way that they can be reduced to several uncoupled first order ordinary differential
equations in the following way. Note that there obviously is a constant solution
$F_0=C$ for the first equation. Inserting it into (\ref{eq:eomFSortedByOrders2})
and (\ref{eq:eomFSortedByOrders4}) leaves us with ordinary differential equations
for $F_{1/2}'$ and $G_1'$ respectively. Using the solutions of $F_0$ and $F_{1/2}$
in (\ref{eq:eomFSortedByOrders3}) gives one more such equation for $F_1'$.

To fix the boundary values of the solutions just mentioned, we demand the value
of $F(u_H=1)$ to be given by the constant $F_0$ and therefore choose the other
component functions' solutions such that $\lim_{u\to 1}F_{1/2}=0$, 
and the same for $F_1$ and $G_1$. The remaining integration constant $C$ is
determined by taking the boundary limit $u \to 0$ of the explicit solution
(\ref{eq:fullX}), making use of the second boundary condition
\begin{equation}
	\lim_{u\to 0} X(u) = X^{\text{bdy}},
\end{equation}
see appendix~\ref{sec:solutionsEOM}.  Eventually, we end up with all the
ingredients needed to construct the gauge field's fluctuations $X(u)$ as in 
(\ref{eq:fullX}).

We solve the equations (\ref{eq:eomXalpha}) with lower sign for 
$\widetilde X_\alpha$ and (\ref{eq:eomX0}) for $X_0'$ and its tilded
partner in exactly the same way as just outlined, only some coefficients of these
differential equations differ. The solution for $X_3'$ is then obtained from
(\ref{eq:eomX3}).

All solutions are given explicitly in Appendix~\ref{sec:solutionsEOM}
together with all other information needed to construct the functions
$X_\alpha$, $\widetilde X_\alpha$, $X_0'$, $\widetilde X_0'$, $X_3'$ and
$\widetilde X_3'$.

\section{Isospin diffusion and correlation functions}
\label{sec:heatAndDiff}

\subsection{Current correlators}
\label{sec:correlators}
In this section we obtain the momentum space correlation functions
for the gauge field component combinations $X$ and
$\widetilde X$ defined in equation~(\ref{eq:flavorTrafo}).
Recall  
that the imaginary part of the retarded correlators essentially gives 
the thermal spectral functions~(see also section~\ref{sec:hydroAdS}). 
The following discussion of the correlators' properties is therefore 
equivalent to a discussion of the corresponding spectral functions.  

First note that
the on-shell action gets new contributions
from the non-Abelian structure,  
\begin{align}      
\label{eq:onShellActionOfA}
S_{{\mathrm D7}} =\;& -T_{7} \frac{(2 \pi \alpha')^2}{2} 2\pi^2 R^3 \, T_r\\
        & \times 2 \int
  \frac{\mathrm d^4 q}{(2 \pi)^4}\: \left[\left. \sqrt{-g}g^{uu}g^{jj'}\:
    {A^a_{j}}'(\vec q) \, 
   A^a_{j'}(-\vec q)   
  \right|_{u_b=0}^{u_h=1}  -
  4 i q \, f^{abc} \mu^c \int\limits_0^1 \mathrm d u \:
  \sqrt{-g}g^{00}g^{33}A_{[3}^{a} A^b_{0]}\,
  \right] \, , \nonumber   
\end{align}
where $j,\,j'\,=\,0\,,1\,,2\,,3$ and the index $u$ denotes the radial
AdS-direction.  
Up to the sum over flavor indices, the first term in the bracket is similar to
the Abelian Super-Maxwell action of \cite{Policastro:2002se}.
The second term is a new contribution depending on the
isospin chemical potential.   
It is a contact term which we will neglect.
The correlation functions however get a structure
that is different from the Abelian case. This is due to the appearance of
the chemical potential in the equations of motion and their solutions. 
Writing~(\ref{eq:onShellActionOfA}) as a function of
$X$ and $\widetilde X$ results in
\begin{align}
\label{eq:onShellActionOfX}
S_{{\mathrm D7}} =\;& -T_{7} \frac{(2 \pi \alpha')^2}{2} 2\pi^2 R^3 \, T_r
        \nonumber  \\
        & \times 2 \int \frac{\mathrm d^4 q}{(2 \pi)^4}\:
        \Big \{\left. \sqrt{-g}\,g^{uu}g^{jj'}\left[
         \frac{1}{2}\left({X_j}'\widetilde{X}_{j'}
         +{\widetilde{X}_j}'{X_{j'}}\right)  +
           {A^3_j}' A^3_{j'}
        \right]\right|_{u_b=0}^{u_h=1}  \\ \nonumber
        &\qquad\qquad\qquad -4  q \mu \int\limits_0^1
         \mathrm d u \sqrt{-g}g^{00}g^{33}
        \left(X_{[0} \widetilde{X}_{3]}+\widetilde{X}_{[3} X_{0]}
         \right) \Big\}  .
\end{align}

In order to find the current correlators, we apply the method outlined in
section~\ref{sec:hydroAdS} to~(\ref{eq:onShellActionOfX}), with the solutions
for the fields given in appendix~\ref{sec:solutionsEOM}.
As an example,
we derive the correlators $G_{0\widetilde 0}=\langle J_0(\vec q)
\widetilde{J}_0(-\vec q)\,\rangle$ and $G_{\widetilde 0 0}=
 \langle \widetilde{J}_0(\vec q)
\,{J_0}(-\vec q)\rangle$ of the flavor current time components 
$J_0$ and $\widetilde{J}_0$,
coupling to the bulk fields $X_0$ and
$\widetilde{X}_0$, respectively. Correlation functions
of all other components are derived analogously. For the notation see appendix
\ref{sec:notation}.

\subsubsection{Green functions: Calculation}

First, we extract the factor~$B(u)$
of~(\ref{eq:classicalAction}),
\begin{equation}
\label{eq:bOfU}
B(u)\,=\, -T_7 \frac{(2\pi\alpha')^2}{2} 2\pi^2 R^3 T_r
    \sqrt{-g}\,g^{uu}\,g^{00} \, .
\end{equation}
The second step, finding the solutions to the mode equations
of motion, has already been performed in section~\ref{sec:solutions}.
In the example at hand we need the
solutions~$X_0$ and~$\widetilde X_0$. From~(\ref{eq:fullX})
and from appendix~\ref{sec:solutionsEOM} we obtain
\begin{align}
\label{eq:derivativesX0}
X_0{}'=&-(1-u)^{\sqrt{\frac{\wn\mn}{2}}}\,
 \frac{\qn^2 \widetilde X_0^{\text{bdy}}+\wn\qn \widetilde X_3^{\text{bdy}}}
  {\sqrt{2\mn\wn}+\wn\mn \ln 2+\qn^2}\,
  \left[
   1-\wn^{1/2}\,\sqrt{\frac{\mn}{2}}
   \ln\left(\frac{2u^2}{u+1}\right) \nonumber \right.\\
   &\left.-\wn\frac{\mn}{12}\left(
   \pi^2+3\ln^2 2+3\ln^2(1+u)+6\ln 2 \ln\left(
    \frac{u^2}{1+u}\right) \right.\right.  \\  \nonumber
   &\left.\left. +12 \mathrm{Li}_2(1-u)+
    12\mathrm{Li}_2(-u)-12\mathrm{Li}_2\left(\frac{1-u}{2}\right)
   \right)+\qn^2\ln\left(\frac{u+1}{2u}\right)
  \right] \, , \\  
\label{eq:derivativesX0t}
\widetilde X_0{}'=&\,\hphantom{-}(1-u)^{-i\sqrt{\frac{\wn\mn}{2}}}\,
  \frac{\qn^2 X_0^{\text{bdy}}+\wn\qn X_3^{\text{bdy}}}
  {i\sqrt{2\mn\wn}+\wn\mn \ln 2-\qn^2}\,
  \left[
   1+\wn^{1/2}\,i\sqrt{\frac{\mn}{2}}
   \ln\left(\frac{2u^2}{u+1}\right) \nonumber \right.\\
   &\left.+\wn\frac{\mn}{12}\left(
   \pi^2+3\ln^2 2+3\ln^2(1+u)+6\ln 2 \ln\left(
    \frac{u^2}{1+u}\right) \right.\right.  \\ \nonumber   
   &\left.\left. +12 \mathrm{Li}_2(1-u)+
    12\mathrm{Li}_2(-u)-12\mathrm{Li}_2\left(\frac{1-u}{2}\right)
   \right)+\qn^2\ln\left(\frac{u+1}{2u}\right)
  \right] . 
\end{align}
Note that we need the derivatives to apply~(\ref{eq:retardedThermalGreen}).

Now we perform the third step and insert~(\ref{eq:bOfU}),~(\ref{eq:derivativesX0})
and~(\ref{eq:derivativesX0t}) into~(\ref{eq:retardedThermalGreen}).
Our solutions $X_0$ and $\widetilde X_0$ replace the
solution~$f(u,\vec k)$ and~$f(u,-\vec k)$ in~(\ref{eq:retardedThermalGreen}).
The resulting expression is evaluated at~$u_b=0$, which
comes from the lower limit of the $u$-integral in the on-shell
action~(\ref{eq:onShellActionOfX}).
At small $u=\epsilon\to 0$, (\ref{eq:derivativesX0})
and~(\ref{eq:derivativesX0t}) give
\begin{align}
\label{eq:smallUderivativesX0}
\lim\limits_{u\to 0} X_0{}'=& -
\frac{\qn^2 \widetilde X_0^{\text{bdy}}+\wn\qn \widetilde X_3^{\text{bdy}}}
  {\sqrt{2\mn\wn}+\wn\mn \ln 2+\qn^2}
  -\lim\limits_{\epsilon\to 0} \left(
 \qn^2 \widetilde X_0^{\text{bdy}}+\wn\qn \widetilde X_3^{\text{bdy}}
\right)\ln \epsilon \, ,  \\
\label{eq:smallUderivativesX0t}
\lim\limits_{u\to 0} \widetilde X_0{}'=  
&\hphantom{-}
\frac{\qn^2 X_0^{\text{bdy}}+\wn\qn X_3^{\text{bdy}}}
  {i\sqrt{2\mn\wn}+\wn\mn \ln 2-\qn^2}
  +\lim\limits_{\epsilon\to 0} \left(
 \qn^2 X_0^{\text{bdy}}+\wn\qn X_3^{\text{bdy}}\right)\, \ln \epsilon \, .
\end{align}
In the next to leading order of~(\ref{eq:smallUderivativesX0})
and~(\ref{eq:smallUderivativesX0t})
there appear singularities, just like in the Abelian
Super-Maxwell calculation~\cite[equation~(5.15)]{Policastro:2002se}.
However, in the hydrodynamic limit, we consider only
the finite leading order.

\subsubsection{Green functions: Results}

Putting everything together,
for the two Green functions for the field components $X_0$, $\widetilde X_0$
given in (\ref{eq:flavorTrafo}) by
\begin{equation*}
X_0 = A_0^1 + i A_0^2,\qquad
\widetilde X_0 =  A_0^1 - i A_0^2,
\end{equation*}
we obtain
\begin{align}
\label{eq:GX0X0Traw}
G_{0\widetilde 0}=&\,
  \frac{N_c T }
  {8\pi}\:\frac{2\pi T\,\qn^2}{
  -\sqrt{2\mn\wn}-\qn^2 -
  \wn\mn\,{\mathrm{ln}2} }\, , \\
\label{eq:GX0TX0raw}
G_{\widetilde 0 0}=&\,
  \frac{N_c T }
  {8\pi} \: \frac{2\pi T\, \qn^2}{
  i\sqrt{2\mn\wn}-\qn^2 +
  \wn\mn\, {\mathrm{ln}2} }\, .
\end{align}
These are the Green functions for the time components in Minkowski space,
perpendicular to the chemical potential in flavor space.
All Green functions are obtained considering hydrodynamic
approximations in $\mathcal{O}(\wn^{1/2},\wn,\qn^2)$, neglecting mixed
and higher orders $\mathcal{O}(\wn^{3/2},\wn^{1/2}\qn^2,\qn^4)$.

\begin{figure}
	\includegraphics[width=.45\linewidth]{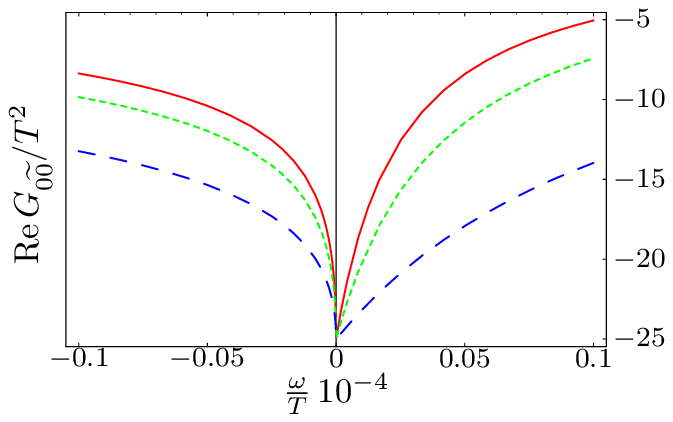}
	\hfill
	\includegraphics[width=.45\linewidth]{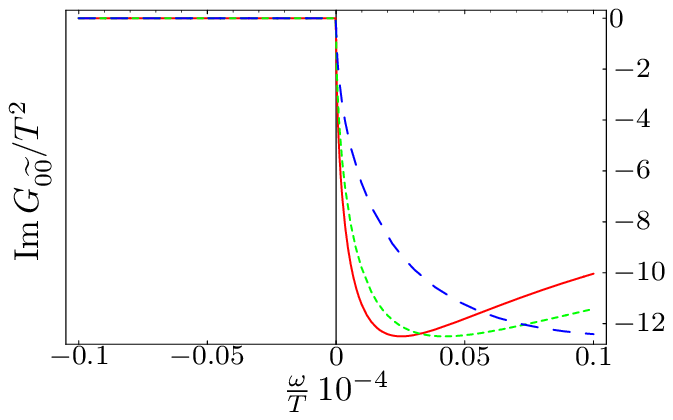}
	\caption{
		Real~(left plot) and imaginary part~(right plot)
		of the correlator $G_{0\widetilde 0}$
		as a function of frequency~$\omega/T$
		at different chemical potential values $\mu/T=0.5$~(solid line),
		$\mu/T=0.3$~(short-dashed line) and $\mu/T=0.1$~(long-dashed line).
		The corresponding plots for the correlator $G_{0\widetilde 0}$ 
		would look like the mirror image of the ones given.
		The real part would be reflected about the vertical
		axis at $\omega=0$, the imaginary part would be
		reflected about the origin.
		All dimensionful quantities are given in units of temperature.
		The numerical values used for the parameters are
		$q/T=0.1$, $N_c=100$.
	}
	\label{fig:reImGX0X0tMu} 
\end{figure}

The prefactor in (\ref{eq:GX0X0Traw}), (\ref{eq:GX0TX0raw}) 
is obtained using $T_7$ as in (\ref{eq:actionDefinitions}), and
carefully inserting all metric factors, together with the standard
AdS/CFT relation $R^4= 4 \pi g_s N_c \alpha'{}^2$. 
As in other
settings with flavor~\cite{Mateos:2006yd},
we concordantly get an overall factor of $N_c$, and not $N_c^2$, for
all correlators. Contrary to those approaches, we do not get
a factor of~$N_f$ when summing over the different flavors.
This is due to the fact that in our setup,  the individual
flavors yield distinct contributions.
Most striking is the non-trivial dependence on the~(dimensionless) chemical
potential~$\mn$ in both correlators.
Note also the distinct
structures in the denominators. The first one~(\ref{eq:GX0X0Traw})
has an explicit relative factor of $i$ between the terms in the denominator.
In the second correlator~(\ref{eq:GX0TX0raw}) there is no explicit
factor of $i$.
The correlator~(\ref{eq:GX0X0Traw})
has a complex pole structure for $\omega>0$,
but is entirely real for $\omega<0$.
On the other hand, (\ref{eq:GX0TX0raw}) is real for
$\omega>0$ but develops a diffusion structure for $\omega<0$. So the
correlators $G_{0 \widetilde 0}$ and $G_{\widetilde 0 0}$
essentially exchange their roles as $\omega$ changes
sign~(see also Fig.~\ref{fig:reImGX0X0tMu}). We find a similar behavior
for all correlators
$G_{j \widetilde l}$ and $G_{\widetilde j l}$
with $j,l\,=\,0,\,1,\,2,\,3$.
This behavior is a consequence of the insertion of $\mathcal{O}(\wn^{1/2})$ in
the hydrodynamic expansion~(\ref{eq:hydroLimit}).

We assume $\mn$ to be small enough in order to neglect
the denominator term of order
$\mathcal{O}(\wn\mn)\ll\mathcal{O}(\sqrt{\wn\mn},\qn^2)$.
Moreover, using the definitions of $\wn,\,\qn$ and $\mn$
from~(\ref{eq:definitionWQM}) we may write
(\ref{eq:GX0X0Traw}) and~(\ref{eq:GX0TX0raw})
as
\begin{align} 
\label{eq:GX0X0T}
G_{0\widetilde 0}=&  
  \,-\frac{N_c T }
  {8\pi\sqrt{2\mu}}\:\frac{q^2 \sqrt{\omega}}{
  \omega+q^2 D(\omega)
  }\, ,\\   
\label{eq:GX0TX0} 
G_{\widetilde 0 0}=&   
  \,\hphantom{-}
  \frac{N_c T }
  {8\pi\sqrt{2\mu}}\:\frac{q^2 \sqrt{\omega}}{
  i\omega-q^2 D(\omega)
  }\, ,  
\end{align}
where the frequency-dependent diffusion coefficient
$D(\omega)$ is given by
\begin{equation} 
\label{eq:diffusionCoeff}
D(\omega)=\sqrt{\frac{\omega}{2\mu}}\:\frac{1}{2\pi T}\, .
\end{equation}
We observe that this coefficient also depends on the
inverse square root of the chemical potential~$\mu$. Its physical
interpretation is discussed below in section \ref{sec:diffusion}.

In the same way we derive the other correlation functions
\begin{align}
\label{eq:GX3X3}
G_{3\widetilde 3}=&
  -\frac{N_c T}
  {8\pi\sqrt{2\mu}}
  \:\frac{\omega^{3/2}\,(\omega-\mu)}{\widetilde{Q}(\omega,q)} \:,&
G_{\widetilde 3 3}=&
  \frac{N_c T}
  {8\pi\sqrt{2\mu}}
  \:\frac{\omega^{3/2}\, (\omega+\mu)}{Q(\omega,q)}\:,& \\
\label{eq:GX0X3}
G_{0\widetilde 3}=&
  -\frac{N_c T}
  {8\pi\sqrt{2\mu}}
  \:\frac{\sqrt{\omega}\,q(\omega  -\mu )}{\widetilde{Q}(\omega,q)}\:,&
G_{\widetilde 0 3}=&
  \frac{N_c T}
  {8\pi\sqrt{2\mu}}
  \:\frac{\sqrt{\omega}\,q(\omega +\mu )}{Q(\omega,q)}\:, & \\
\label{eq:GX3X0}
G_{3\widetilde 0}=&
  -\frac{N_c T}
  {8\pi\sqrt{2\mu}}
  \:\frac{\omega^{3/2}\,q}{\widetilde{Q}(\omega,q)}\: ,  &
G_{\widetilde 3 0}=&
  \frac{N_c T}
  {8\pi\sqrt{2\mu}}
  \:\frac{\omega^{3/2}\,q}{Q(\omega,q)} \:.&
\end{align}
with the short-hand notation 
\begin{equation}
\label{eq:denominatorQs}
Q(\omega,q)= i\omega-q^2 D(\omega), \qquad
\widetilde Q(\omega,q) = \omega+q^2 D(\omega) \, .
\end{equation}
Note that most of these functions are proportional to powers of
$q$ and therefore vanish in the limit of vanishing spatial
momentum $q\to 0$. Only the \mbox{$33$-combinations} from~(\ref{eq:GX3X3})
survive this limit. In contrast to the Abelian Super-Maxwell
correlators~\cite{Policastro:2002se} given
in appendix~\ref{sec:abelianCorrelators}, it stands out that our
results~(\ref{eq:GX0X0T}), (\ref{eq:GX0TX0})
and~(\ref{eq:GX3X3}) and~(\ref{eq:GX3X0}) have
a new zero at $\omega =\,\mu$ or $-\mu$.
Nevertheless bear in mind that we took the limit
$\omega < \mu$ in order to obtain our solutions. Therefore the
apparent zeros at $\pm \mu$ lie outside of the range considered.
Compared to the Abelian case there is an additional factor of $\sqrt{\omega}$.
The dependence on temperature remains linear.
  
In the remaining $X$-correlators we do not find any pole structure
to order $\sqrt{\omega}$, subtracting an $\mathcal{O}(q^2)$ contribution as in
\cite{Policastro:2002se},
\begin{align}
\label{eq:GXalphaXalphaT}
G_{1 \widetilde 1}=& G_{2 \widetilde 2}=
\frac{\sqrt 2 N_c T}{8\pi}\sqrt{\mu\omega}\, , \\
\label{eq:GXalphaTXalpha}
G_{\widetilde 1 1}=& G_{\widetilde 2 2}=
-\frac{i\sqrt 2 N_c T}{8\pi}\sqrt{\mu\omega}\, .
\end{align}
As seen from~(\ref{eq:GXalphaXalphaT}), $G_{\alpha \widetilde \alpha}$
(with $\alpha=1,2$)
are real for negative $\omega$ and imaginary for positive 
$\omega$. The opposite is true for $G_{\widetilde \alpha \alpha}$, as
is obvious from the relative factor of $i$.
  
The correlators of components, pointing along
the isospin potential in flavor space~($a=3$), are found to be
\begin{align} 
\label{eq:GA03A03}
&G_{A_0^3A_0^3}=  
  \frac{N_c T}{4 \pi}\:\frac{q^2}{i\omega-D_0 q^2}\: ,&
&G_{A_0^3A_3^3}=G_{A_3^3A_0^3}=
  \frac{N_c T}{4 \pi}\:\frac{\omega q}{i\omega-D_0 q^2} \, ,&  \\
\label{eq:GAalpha3Aalpha3}
&G_{A_1^3A_1^3}=G_{A_2^3A_2^3}=-\frac{N_c T\,i\omega}{4 \pi}\: ,&
&G_{A_3^3A_3^3}=\frac{N_c T}{4 \pi}\:\frac{\omega^2}{i\omega-D_0 q^2} \, ,&
\end{align}    
with the diffusion constant~$D_0=1/(2\pi T)$ .
Note that these latter correlators have the same structure 
but differ by a factor~$4/N_c$ from those found in the 
Abelian super-Maxwell case~\cite{Policastro:2002se}~(see 
also~(\ref{eq:abelianCorrelators1}) and~(\ref{eq:abelianCorrelators2})). 
In particular the correlators in equation~(\ref{eq:GA03A03}) do not 
depend on the chemical potential.

To analyze the novel structures appearing in the other correlators,
we explore their real and imaginary parts  as well as the interrelations
among them,
\begin{align}
  \label{eq:reGX0X0t>=}
 \mathrm{Re}\,G_{0 \widetilde 0}(\omega\ge 0) & = &
  \hphantom{-\;}
  \mathrm{Re}\,G_{\widetilde 0 0}(\omega < 0) & = &&
  -\frac{N_c T}{8\pi}
  \:\frac{q^2}{
  \sqrt{2\mu\left|\omega\right|}+q^2/(2\pi T) }\, ,\\
\label{eq:reGX0X0t<}
  \mathrm{Re}\,G_{0 \widetilde 0}(\omega < 0) & = &
  \hphantom{-\;}
  \mathrm{Re}\,G_{\widetilde 0 0}(\omega \ge 0) & = &&
  -\frac{N_c T}{16\pi^2} \frac{q^4}{
  2\mu\left|\omega\right|+q^4/(2\pi T)^2 }\,, \\
\label{eq:imGX0X0t>=}
 \mathrm{Im}\,G_{0 \widetilde 0}(\omega< 0) & = &
  -\mathrm{Im}\,G_{\widetilde 0 0}(\omega \ge 0) & = &&
  \hphantom{-\:\:}\frac{N_c T}{8\pi}\:\frac{q^2 \sqrt{2\mu\left|\omega\right|}}
   {
  2\mu\left|\omega\right|+q^4/(2\pi T)^2 }\, , 
\end{align}
and
\begin{equation}
	\label{eq:imGX0X0t<}
  \mathrm{Im}\,G_{0 \widetilde 0}(\omega \ge 0) = 0,\qquad
  \mathrm{Im}\,G_{\widetilde 0 0}(\omega  <  0) = 0.
\end{equation}
Now we see why, as discussed below~(\ref{eq:GX0TX0}),
$G_{0 \widetilde 0}$ and $G_{\widetilde 0 0}$ exchange
their roles when crossing the origin  at $\omega=0$.
This is
due to the fact that the real parts
of all $G_{j\widetilde l}$ and $G_{\widetilde j l}$ are
mirror images of each other by reflection about the vertical axis
at $\omega=0$. In contrast, the imaginary parts are inverted into
each other at the origin.
Figure~\ref{fig:reImGX0X0tMu} shows the real and
imaginary parts of correlators
$G_{0 \widetilde 0}$ and $G_{\widetilde 0 0}$.
The different curves correspond to distinct values
of the chemical potential~$\mu$. The real part shows a deformed resonance
behavior. The imaginary part has a deformed interference shape with
vanishing value for negative frequencies.
All curves are continuous and finite at
$\omega = 0$.
However due to the square root dependence, they are not differentiable
at the origin. Parts of the correlator which are real
for positive $\omega$ are shifted into the imaginary part by the
change of sign when crossing $\omega=0$, and vice versa.

To obtain physically meaningful correlators, we follow a procedure
which generalizes the Abelian approach of~\cite{Kovtun:2005ev}.
In the Abelian case, gauge-invariant components of the field strength
tensor, such as  $E_\alpha = \omega A_\alpha$, are
considered as physical variables. This procedure cannot be transferred
directly to the non-Abelian case. Instead, we consider the non-local
part of the gauge invariant ${\rm tr}  F^2$ which contributes to the
on-shell action (\ref{eq:onShellActionOfA}). In this action, the
contribution involving the non-Abelian structure constant -- as well
as  $\mu$ -- is a
local contact term. The non-local contribution however generates 
the Green function combination
\begin{equation}
\label{eq:sumOfGA0A0}
G_{A_i^1 A_j^1}+G_{A_i^2 A_j^2}+G_{A^3_i A^3_j}\, .
\end{equation}
We take this sum as our physical Green function. This choice is
supported further by the fact that it may be written in terms
of the linear combinations (\ref{eq:flavorTrafo}) 
which decouple the equations of motion.
For example, for the time component, written in the variables~
$X_0,\,\widetilde X_0$ given by~(\ref{eq:flavorTrafo}),
the combination (\ref{eq:sumOfGA0A0}) 
reads~(compare to~(\ref{eq:onShellActionOfX}))
\begin{equation}
\label{eq:sumOfGX0X0} 
G_{0\widetilde 0}+G_{\widetilde 0 0}+G_{A^3_0 A^3_0}\,  .
\end{equation}   
The contribution from~$G_{A^3_0 A^3_0}$ is of order~$\mathcal{O}(\mu^0)$,
while the combination for 
the first two flavor directions, $G_{0\widetilde 0}+G_{\widetilde 0 0}$,
is of order~$\mathcal{O}(\mu)$.

\begin{figure}
	\includegraphics[width=.45\linewidth]{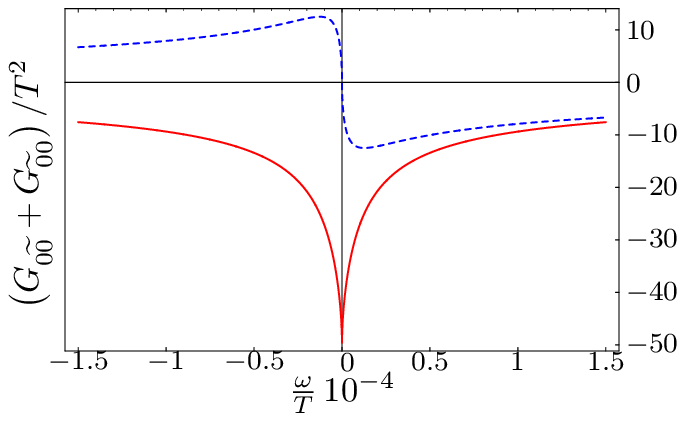}
	\hfill
	\includegraphics[width=.45\linewidth]{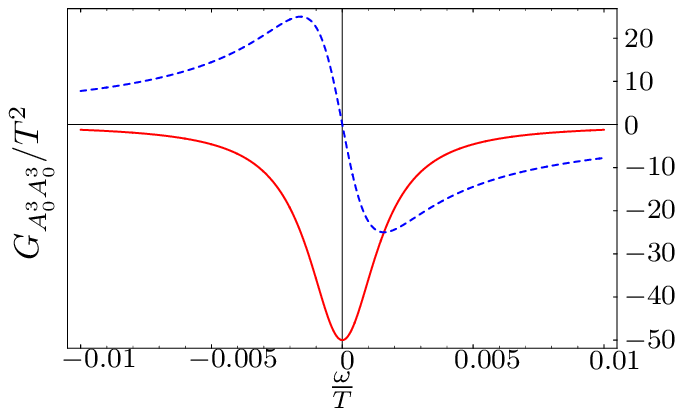}
	\caption{ 
		In the left plot the sum of both correlators in
		$0 0$-directions is split into
		its imaginary~(dashed line) and real~(solid
		line) part and plotted against frequency.
		For comparison the right plot shows the corresponding real and
		imaginary parts for the~$G_{A_0^3A_0^3}$. It is qualitatively
		similar to the Abelian correlator in $i=0$ Lorentz direction computed
		from the Super-Maxwell action in~\cite{Policastro:2002se}.
		Note the different frequency scales in the two plots. The curves
		in~$a=1,\,2$-directions are much narrower due to their square root dependence
		on $\omega$. Furthermore they have a much larger maximum amplitude.
		All dimensionful quantities are given in units of temperature.
		The numerical values used for the parameters are, as in
		Fig.~\ref{fig:reImGX0X0tMu},
		$q/T=0.1$, $N_c=100$ and only in the left plot $\mu/T=0.2$.
	}
	\label{fig:reImGX0X0sum}
\end{figure}

We proceed by discussing the physical behavior of the Green function
combinations introduced above.                  
$G_{A^3_0 A^3_0}$ is plotted in Fig.~\ref{fig:reImGX0X0sum} on the right.
Its frequency dependence is of the same form as in the Abelian
correlator obtained in \cite{Policastro:2002se}, 
as can be seen from~(\ref{eq:abelianCorrelators1}).
Since we are interested in effects of 
order~$\mathcal{O}(\mu)$, we drop the third flavor direction~$a=3$
from the sum~(\ref{eq:sumOfGX0X0}) in
the following.
It is reassuring to observe that the flavor directions~$a=1,\,2$ which are
orthogonal to
the chemical potential, combine to give a correlator spectrum 
qualitatively similar
to the one found in \cite{Policastro:2002se} for the Abelian
Super-Maxwell action~(see Fig.~\ref{fig:reImGX0X0sum}).
However, we discover intriguing new effects
such as the highly increased steepness of the curves near the origin
due to the square root
dependence and a kink at the origin.

We observe a narrowing of the inverse resonance peak compared to the
form found for the Abelian Super-Maxwell action~(and also compared 
to the form of our~$G_{A_0^3A_0^3}$, as is seen from
comparing the left with the right plot in Fig.~\ref{fig:reImGX0X0sum}).
At the origin the real and imaginary part are finite and continuous,
but they are not continuously differentiable. However,
the imaginary part of~$G_{A_0^3A_0^3}$ has finite steepness at the origin.
The real part though  has vanishing derivative at $\omega=0$.
Note that the imaginary part of flavor directions~$a=1,\,2$ on the left
plot in Fig.~\ref{fig:reImGX0X0sum}
never drops below the real part. In the third flavor direction, as well as
in the Abelian solution, such a drop
occurs on the positive $\omega$-axis.

The correlators  $G_{3 \widetilde 3}$, $G_{\widetilde 3 3}$, $G_{0 \widetilde 3}$
and $G_{\widetilde 0 3}$ have the same interrelations between
their respective real and imaginary parts as $G_{0 \widetilde 0}$
and $G_{\widetilde 0 0}$. Nevertheless, their
dependence on the frequency and momentum is different, as
can be seen from~(\ref{eq:GX3X3}) to~(\ref{eq:GX3X0}). A list
of the $33$-direction Green functions split into real and imaginary parts can be
found in appendix~\ref{sec:correlationFunctions}.

Thermal spectral functions in different directions are 
compared graphically in appendix~\ref{sec:thermalSpectralFunctions}. 

\subsection{Isospin diffusion}
\label{sec:diffusion}
The attenuated poles in hydrodynamic correlation functions have
specific meanings~(for exemplary discussions of this in the AdS/CFT setup
see e.g.~\cite{Kovtun:2006pf}, \cite{Policastro:2002tn}).
In our case we observe an attenuated pole
in the sum
$G_{0\widetilde 0}+G_{\widetilde 0 0}$
at $\omega = 0$. As can be seen from the plots in
Fig.~\ref{fig:reImGX0X0sum}, our pole lies at $\mathrm{Re}\,\omega =0$.
This structure appears
in hydrodynamics as the signature of a diffusion pole
located at purely imaginary $\omega$.
Its location on the imaginary $\omega$-axis is given by the
zeros of the denominators of our correlators as~(neglecting
$\mathcal{O}(\omega\,,q^4\,)$)
\begin{equation} 
\label{eq:poleLocation12}
\sqrt{\omega}\,=\,-i\, \frac{ q^2}{2\pi T \sqrt{2\mu}}\,.     
\end{equation}   
Squaring both sides of~(\ref{eq:poleLocation12}) we see that
this effect is of order
$\mathcal{O}(q^4)$. On the other hand, looking for poles in the correlator
involving the third flavor direction~$G_{A_0^3 A_0^3}$, we obtain
dominant contributions of order~$\mathcal{O}(q^2)$
and~$\mathcal{O}(\mu^0)$~(neglecting
$\mathcal{O}(\omega^2\,,q^2\,)$)
\begin{equation}
\label{eq:poleLocation3}
\omega\,=\,-i\, \frac{ q^2}{2\pi T } \,  .
\end{equation}
This diffusion pole is reminiscent of the result of the Abelian result
of \cite{Policastro:2002se}
given in appendix~(\ref{sec:abelianCorrelators}).
As discussed in section~\ref{sec:correlators},
we consider gauge-invariant
combinations~$G_{0\widetilde 0}+G_{\widetilde 0 0}+G_{A_0^3 A_0^3}$.
In order
to inspect the non-Abelian effects of order~$\mathcal{O}(\mu)$ showing up in the
first two correlators in this sum, we again drop the third
flavor direction~which is of order~$\mathcal{O}(\mu^0)$.

Motivated by the diffusion pole behavior of our correlators in
flavor-directions~$a=1,\,2$ corresponding to the combinations
$X,\, \widetilde X$~(see~(\ref{eq:poleLocation12})),
we wish to regain the structure of
the diffusion equation given in~(\ref{eq:omegaD}), which
in our coordinates~($k\,=\,(\omega\,,0\,,0\,,q\,)$) reads
\begin{equation}
\label{eq:diffusionEquationFT}
i\,\omega\, J_0=\,D(\omega)\, q^2\,J_0 \, .
\end{equation}
Our goal is to rewrite~(\ref{eq:poleLocation12}) such that
a term~$\mathcal{O}(\omega)$ and one term in
order~$\mathcal{O}(q^2)$ appears. Furthermore there should
be a relative factor of $-i$ between these two terms.
The obvious manipulation to meet these requirements is
to multiply~(\ref{eq:poleLocation12}) by $\sqrt{\omega}$ in order
to get
\begin{equation}
\label{eq:poleLocation12timesSqrtW}
\omega\,=\,-i\,q^2\, \frac{\sqrt{\omega}}{2\pi
 T \sqrt{2\mu}} \,  .
\end{equation}
Comparing the gravity result~(\ref{eq:poleLocation12timesSqrtW})
with the hydrodynamic equation~(\ref{eq:diffusionEquationFT}), 
we obtain the frequency-dependent diffusion coefficient
\begin{equation}
\label{eq:diffusionCoefficient2}
D(\omega)\,=\, \sqrt{\frac{\omega}{2\,\mu}}
  \frac{1}{2\pi\,T} \,  .
\end{equation}
Our argument is thus summarized as follows:
Given the isospin chemical potential as in~(\ref{eq:backg}),
(\ref{eq:backgroundRule}),
$J_0$ from (\ref{eq:omegaD}) is the isospin charge
density in~(\ref{eq:diffusionEquationFT}). According to
(\ref{eq:diffusionEquationFT}), 
the coefficient~(\ref{eq:diffusionCoefficient2})
describes the diffusive  response of the
quark-gluon plasma to a gradient in the isospin charge
distribution. For this reason we interprete $D(\omega)$ as the isospin 
diffusion coefficient.

Near the pole, the strongly-coupled plasma behaves analogously
to a diffractive medium with anomalous dispersion
in optics. In the presence of the isospin chemical potential, the propagation
of non-Abelian gauge fields in the black hole background depends on the square
root of the frequency. In the dual gauge theory, this corresponds to a
non-exponential decay of isospin fluctuations with time. 

The square root dependence of our diffusion 
coefficient is valid for small frequencies. As long as 
$\omega/T<1/4$, the square root is larger than its argument and 
at $\omega/T=1/4$, the difference to a linear dependence on 
frequency is maximal. 
Therefore in the regime of small frequencies~$\omega/T<1/4$, which is 
accessible to our approximation, diffusion of modes close to $1/4$ is enhanced
compared to modes with frequencies close to zero.

\section{Conclusion}
\label{sec:conclusion}
In this paper we have considered a relatively simple gauge/gravity dual model
for a finite temperature field theory, 
consisting of an isospin chemical 
potential $\mu$ obtained from a time component vev for the $SU(2)$ gauge
field on two coincident brane probes. We have considered the D7-brane
embedding corresponding to vanishing quark mass, for which $\mu$ is a
constant, independent in particular of the radial holographic coordinate.
The main result of this paper is that this model, despite its simplicity, 
leads to a hydrodynamical behavior of the dual field theory which goes 
beyond linear response theory. We find in particular a frequency-dependent
diffusion coefficient with a non-analytical behavior.
Frequency-dependent diffusion is a well-known
phenomenon in condensed matter physics. Here it originates simply from the
fact that due to the non-Abelian structure of the gauge field on the 
brane probe, the chemical potential replaces a time derivative in the action
and in the equations of motion from which the Green functions are obtained. 

Of course the calculation presented has some limitations as far as the
approximations made are concerned. This applies in particular to 
the approximation (\ref{eq:approximateIndices}) 
of the so-called indices in the ansatz for solving
the equations of motion. Here we have dropped the constant present
under the square root and used an expression proportional to the
square root of the frequency. This allows for a closed solution
without having to use numerics. However using this approximation we
have dropped the Abelian limit. This leads ultimately to the 
square root dependence of the diffusion coefficient on the frequency. 
This dependence is unphysical for $\omega \rightarrow 0$, since the
diffusion coefficient is expected to be non-zero for zero frequency. 
We expect physical behavior to be restored if the Abelian limit is included 
in the calculation. To avoid the approximation described, this
requires a numerical approach. A suitable solution method for all
momenta and all frequencies has been presented in 
\cite{Kovtun:2006pf} and in \cite{Teaney:2006nc}.  
We are going to study the application of this method to the model
presented here in the future.

\begin{acknowledgments}
\label{sec:ack}
We are grateful to R.~Apreda, 
G.~Policastro, C.~Sieg, A.~Starinets and  L.~Yaffe
for useful discussions and correspondence.
\end{acknowledgments}

\begin{appendix}

\section{Notation}
\label{sec:notation}
The five-dimensional $AdS$ Schwarzschild black hole space in which we
work is  endowed with a metric of
signature $(-,+,+,+,+)$, as  given explicitly in (\ref{eq:finiteTMetric}).
We make use of the Einstein notation to indicate sums over Lorentz indices,
and additionally simply sum over non-Lorentz indices, such as
 gauge group indices, whenever they occur twice in a term.

To distinguish between vectors in different dimensions of the $AdS$ space, we
use bold symbols like $\bm{q}$ for vectors in the the \emph{three spatial
dimensions} which do not live along the radial $AdS$ coordinate.
\emph{Four-vectors} which do not have components along the radial $AdS$
coordinates are denoted by symbols with an arrow on top, as $\vec{q}$.

The Green functions $G=\langle J \hat J \rangle$ considered give correlations
between currents $J$ and $\hat J$. These currents couple to fields $A$ and
$\hat A$ respectively. In our notation
we use symbols such as $G_{A^a_k A^b_l}$ to denote correlators of currents
coupling to fields $A^a_k$ and $A^b_l$, with flavor indices $a,b$ and Lorentz
indices $k,l=0,1,2,3$. For the gauge field combinations $X_k$ and $\widetilde X_l$
given in (\ref{eq:flavorTrafo}), we obtain Green functions $G_{k\widetilde l}$
 denoting correlators of the corresponding currents.

\section{Solutions to equations of motion}
\label{sec:solutionsEOM}

Here we explicitly write down the component functions used to construct the
solutions to the equations of motion for the gauge field fluctuations up to
order $\wn$ and $\qn^2$. The functions themselves are then composed as in
(\ref{eq:fullX}).

The solutions for the components with flavor index
$a=3$ where obtained in \cite{Policastro:2002se}.

\subsection{Solutions for $X_\alpha$, $\widetilde X_\alpha$ and $A^3_\alpha$}
\label{sec:solXalpha}

The function $X_\alpha(u)$ solves (\ref{eq:eomXalpha}) with the upper sign
and is constructed as in (\ref{eq:fullX}) from the following component functions,
\begin{align}
        \label{eq:XalphaBeta}
        \beta   &= \sqrt{\frac{\wn\,\mn}{2}}+\mathcal{O}(\omega),\\[\bigskipamount]
        F_0      &= C,\\
        F_{1/2}  &= - C \sqrt{\frac{\mn}{2}}\;\ln\frac{1+u}{2},\\
        F_1      &=               - C \frac{\mn}{12}\, \Bigg[ \pi^2-9 \ln^22  + 3 \ln(1-u)\left( \ln 16 - 4 \ln(1+u)\right)\nonumber\\
                     &\hphantom{=\,\: - C \frac{\mn}{12}\, \Big[ \pi^2-9 \ln^22} + 3 \ln(1+u) \left( \ln(4(1+u)) -4\ln u \right)\\
                     &\hphantom{=\,\: - C \frac{\mn}{12}\, \Big[ \pi^2-9 \ln^22} - 12\left( \mathrm{Li}_2(1-u)+\mathrm{Li}_2(-u) + \mathrm{Li}_2\left(\frac{1+u}{2}\right)\right) \Bigg]\nonumber,\\
        G_1      &=\frac{C}{2}\,\left[ \frac{\pi^2}{12}+ \ln u\ln(1+u) + \mathrm{Li}_2(1-u)+\mathrm{Li}_2(-u) \right],\\
        \intertext{%
        where the constant $C$ can be expressed it in terms of the field's boundary
        value $X^{\text{bdy}} = \lim_{u\to 0} X(u,k)$,%
        }
        \label{eq:XalphaC}
        C        &= X^{\text{bdy}} \left(1 + \sqrt{\frac{\mn\,\wn}{2}} \ln 2 + \mn\,\wn\left(\frac{\pi^2}{6} + \frac{\ln^2 2}{4}\right) + \frac{\pi^2}{8} \qn^2 + \mathcal{O}(\wn^{3/2},\qn^4) \right)^{-1}.
\end{align}

The solutions of the equations of motion (\ref{eq:eomXalpha}) with lower sign
for the functions $\tilde X_\alpha(u)$ are given by
\begin{align}
        \tilde \beta   &=-i\sqrt{\frac{\wn\,\mn}{2}}+\mathcal{O}(\omega),\\[\bigskipamount]
        \tilde F_0     &= \tilde C,\\
        \tilde F_{1/2} &= i \tilde C \sqrt{\frac{\mn}{2}}\;\ln\frac{1+u}{2},\\
        \tilde F_1     &=               \tilde C \frac{\mn}{12}\, \Bigg[ \pi^2-9 \ln^22  + 3 \ln(1-u)\left( \ln 16 - 4 \ln(1+u)\right)\nonumber\\
                       &\hphantom{=\,\: \tilde C \frac{\mn}{12}\, \Big[ \pi^2-9 \ln^22} + 3 \ln(1+u) \left( \ln(4(1+u)) -4\ln u \right)\\
                       &\hphantom{=\,\: \tilde C \frac{\mn}{12}\, \Big[ \pi^2-9 \ln^22} - 12\left( \mathrm{Li}_2(1-u)+\mathrm{Li}_2(-u) + \mathrm{Li}_2\left(\frac{1+u}{2}\right)\right) \Bigg],\nonumber\\
        \tilde G_1     &=\frac{\tilde C}{2}\,\left[ \frac{\pi^2}{12}+ \ln u\ln(1+u) + \mathrm{Li}_2(1-u)+\mathrm{Li}_2(-u) \right],\\
        \intertext{with $\tilde C$ given by}
        \tilde C        &= \tilde X^{\text{bdy}} \left(1-i\sqrt{\frac{\mn\,\wn}{2}} \ln 2 - \mn\,\wn\left(\frac{\pi^2}{6} + \frac{\ln^2 2}{4}\right) + \frac{\pi^2}{8} \qn^2 + \mathcal{O}(\wn^{3/2},\qn^4) \right)^{-1},
\end{align}
so that $\lim_{u\to 0} \tilde X(u,k)=\tilde X^{\text{bdy}}$.

The solution for $A^3_\alpha$ solves (\ref{eq:eomA1A2Flavor3}) up to order
$\wn$ and $\qn^2$ with boundary value $\left(A^3_\alpha\right)^{\text{bdy}}$. It
is
\begin{equation}
\begin{aligned}
        A^3_\alpha = \frac{8\;\left(A^3_\alpha\right)^{\text{bdy}} (1-u)^{-\frac{i\wn}{2}}}{8-4i\wn\ln 2 + \pi^2\qn^2}
        \Bigg[ \,&  1 + i\frac{\wn}{2} \ln\frac{1+u}{2}\\
                 &  + \frac{\qn^2}{2} \left( \frac{\pi^2}{12}+ \ln u\ln(1+u) + \mathrm{Li}_2(1-u)+\mathrm{Li}_2(-u) \right)\Bigg].
\end{aligned}
\end{equation}

\subsection{Solutions for $X_0'$, $\widetilde{X}_0'$ and ${A^3_0}'$}

Here we state the solutions to (\ref{eq:eomA0}). This formula describes three
equations, differing in the choice of $a=1,2,3$.  The cases $a=1,2$ give
coupled equations which are decoupled by transformation from $A^{1,2}_0$ to
$X_0$ and $\widetilde{X}_0$. The choice $a=3$ gives a single equation.

The function $X_0'$ is solution to (\ref{eq:eomX0}) with upper sign. We specify
the component functions as
\begin{align}
        \beta    &= \sqrt{\frac{\wn\,\mn}{2}}+\mathcal{O}(\omega),\\[\bigskipamount]
        F_0      &= C,\\
        F_{1/2}  &= - C \sqrt{\frac{\mn}{2}}\;\ln\frac{2u^2}{1+u},\\
        F_1      &= - C \frac{\mn}{12}\, \Bigg[ \pi^2 + 3 \ln^22  + 3 \ln^2(1+u) +6 \ln 2 \ln \frac{u^2}{1+u}\\
                                          &\hphantom{= -C \frac{\mn}{12}\, \Bigg[\, } + 12 \left( \mathrm{Li}_2(1-u) +\mathrm{Li}_2(-u) - \mathrm{Li}_2\left(\frac{1-u}{2}\right) \right)\Bigg],\\
        G_1      &= C \ln\frac{1+u}{2u},\\
        \intertext{%
        where the constant $C$ can be expressed in terms of the field's boundary
        value $X^{\text{bdy}} = \lim_{u\to 0} X(u,k)$,%
        }
        C        &= - \frac{\qn^2  X_0^{\text{bdy}} + \wn\qn X_3^{\text{bdy}}}{\sqrt{2\mn\wn} + \mn\wn \ln 2 + \qn^2}.
\end{align}

To get the function $\widetilde{X}_0'$, we solve (\ref{eq:eomX0}) with the lower
sign and obtain
\begin{align}
        \widetilde{\beta}    &= -i\sqrt{\frac{\wn\,\mn}{2}}+\mathcal{O}(\omega),\\[\bigskipamount]
        \widetilde{F}_0      &= \widetilde C,\\
        \widetilde{F}_{1/2}  &= i \widetilde{C} \sqrt{\frac{\mn}{2}}\;\ln\frac{2u^2}{1+u},\\
        \widetilde{F}_1      &          = \widetilde{C} \frac{\mn}{12}\, \Bigg[ \pi^2 + 3 \ln^22  + 3 \ln^2(1+u) +6 \ln 2 \ln \frac{u^2}{1+u}\\
                                          &\hphantom{= \widetilde{C} \frac{\mn}{12}\, \Bigg[\, } + 12 \left( \mathrm{Li}_2(1-u) +\mathrm{Li}_2(-u) - \mathrm{Li}_2\left(\frac{1-u}{2}\right) \right)\Bigg],\\
        \widetilde{G}_1      &=\widetilde{C}\ln\frac{1+u}{2u},\\
        \intertext{%
        where the constant $\widetilde{C}$ can be expressed it in terms of the field's boundary
        value $\widetilde{X}^{\text{bdy}} = \lim_{u\to 0} \widetilde{X}(u,k)$,%
        }
        \widetilde{C} &= \frac{\qn^2  \widetilde{X}_0^{\text{bdy}} + \wn\qn \widetilde{X}_3^{\text{bdy}}}{i \sqrt{2\mn\wn} + \mn\wn \ln 2-\qn^2}.
\end{align}

The solution for (\ref{eq:eomA0}) with $a=3$ is the function ${A^3_0}'$,
given by
\begin{equation}
        {A^3_0}' = (1-u)^{-\frac{i\wn}{2}}\, \frac{\qn^2 A_0^{\text{bdy}}+\wn\qn A_3^{\text{bdy}}}{i\wn-\qn^2} \left(1 + \frac{i\wn}{2}\ln \frac{2u^2}{1+u} + \qn^2 \ln \frac{1+u}{2u}\right).
\end{equation}

\subsection{Solutions for $X_3'$, $\widetilde{X}_3'$ and ${A^3_3}'$}

We give the derivatives of $X_3$ and $\widetilde X_3$ as
\begin{align}
                     X_3' & = -\frac{\wn - \mn}{\qn f}\, X_0'\\
     \widetilde{X}_3' & = -\frac{\wn + \mn}{\qn f}\, \widetilde{X}_0'.
\end{align}

The solution for ${A^3_3}'$ is
\begin{equation}
        {A^3_3}' = -\frac{\wn}{\qn} \, {A^3_0}'.
\end{equation}

\subsection{Comparison of numerical and analytical results}
\label{sec:compareAnaNum}
\begin{figure}
	\includegraphics[width=.45\linewidth]{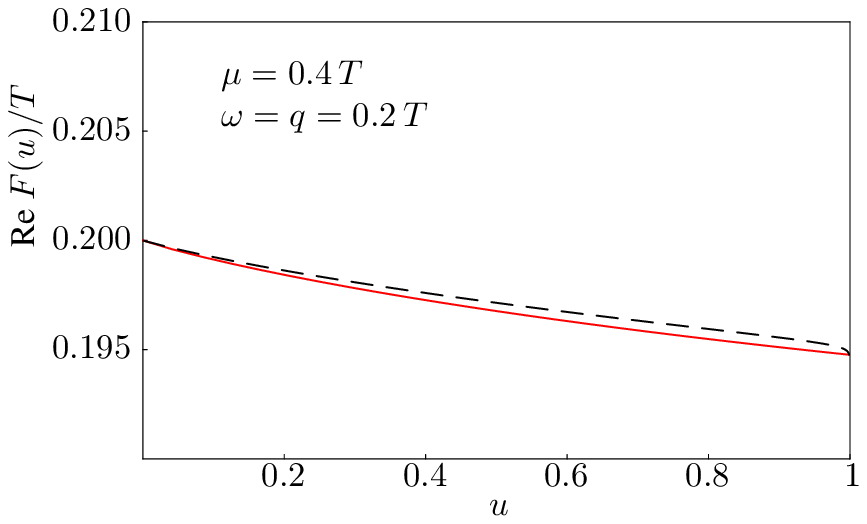}
	\hfill
	\includegraphics[width=.45\linewidth]{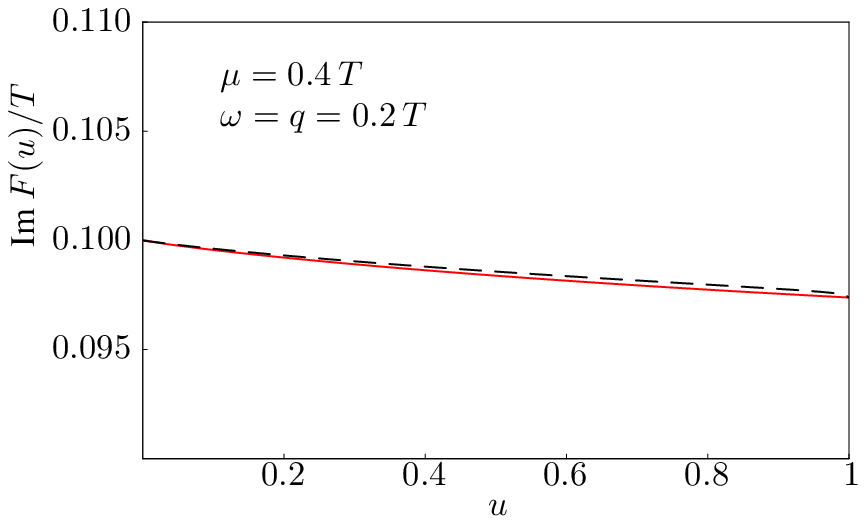}
	\caption{These plots show the real and imaginary part of the function
			$F(u)$ which is part of $X_\alpha=(1-u)^\beta F(u)$. The solid line
			depicts the analytical approximation, obtained in this paper. As a check
			we solved the equations of motion for $F(u)$ numerically. They
			are drawn as dashed lines. In this example we used $T=1$. The numerical
			solution was chosen to agree with the analytical one at the horizon and
			boundary.
	}
	\label{fig:compareAnaNum}
\end{figure}

As an example, in Fig.~\ref{fig:compareAnaNum} we show the numerical and
analytical solutions for the
function $F(u)$ in $X_\alpha=(1-u)^\beta F(u)$. Here we compare the numerical
result for $F(u)$ obtained from the ansatz (\ref{eq:ansatzA1A2}) with
(\ref{eq:indices}) in (\ref{eq:eomXalpha}) with the analytically obtained
approximation given above in (\ref{eq:XalphaBeta}) to (\ref{eq:XalphaC}). 

\section{Abelian Correlators}
\label{sec:abelianCorrelators}
For convenient reference we quote here the correlation functions 
of the Abelian super-Maxwell theory found in~\cite{Policastro:2002se}. 
The authors start from a 5-dimensional supergravity action 
and not from a Dirac-Born-Infeld action as we do. Therefore there 
is generally a difference by a factor~$N_c/4$. Note also that 
here all~$N_f$ flavors contribute equally. In our notation   
\begin{align}  
\label{eq:abelianCorrelators1}  
&G_{11}^{a b} = G_{22}^{a b}  =- \frac{i N_c^2 T\omega\; \delta^{a b}
}{16\pi}
 + \cdots
\,,\quad  & 
&G_{00}^{a b} = \frac{ N_c^2 T q^2\; \delta^{a b}}{ 16 \pi  ( i\omega - D q^2)}
 + \cdots
\,,\\
\label{eq:abelianCorrelators2}  
&G_{0 3}^{a b} = G_{3 0}^{a b}  = -
  \frac{ N_c^2 T \omega q\; \delta^{a b}}{ 16 \pi  ( i\omega - D q^2)}
 + \cdots
\,,\quad   &  
&G_{33}^{a b} =  \frac{ N_c^2 T \omega^2\;
  \delta^{a b}}{ 16 \pi  ( i\omega - D q^2)}
 + \cdots,  
\end{align}  
where $D=1/(2\pi T)$ .  
\section{Correlation functions}
\label{sec:correlationFunctions}

In this appendix we list the real and imaginary parts of the 
flavor currents in the first two flavor-directions~$a=1,\,2$ and 
in the third Lorentz-direction coupling to the 
supergravity-fields~$X_3$ 
and~$\widetilde X_3$~(as defined in~(\ref{eq:flavorTrafo})).   
\begin{align}
\label{eq:reGX3X3t>=}
\mathrm{Re}\{G_{3 \widetilde 3}(\omega\ge 0)\} & = &
\hphantom{--}
\mathrm{Re}\{G_{\widetilde 3 3}(\omega < 0)\}  & = &&
-\frac{N_c\, q^2\, (\omega^2+\mu\left|\omega\right|)}{16\pi^2 \left[
2\mu\left|\omega\right|+q^4/(2\pi T)^2 \right]}\, , \\
\label{eq:imGX3X3t>=}
\mathrm{Im}\{G_{3 \widetilde 3}(\omega\ge 0)\} & = &
-\mathrm{Im}\{G_{\widetilde 3 3}(\omega < 0)\} & = &&
-\frac{N_c T\, \sqrt{2\mu\left|\omega\right|}\,
 (\omega^2+\mu\left|\omega\right|)}
 {8\pi \left[
2\mu\left|\omega\right|+q^4/(2\pi T)^2 \right]}\, ,\\
\label{eq:reGX3X3t<}
\mathrm{Re}\{G_{3 \widetilde 3}(\omega < 0)\}   & = &
\mathrm{Re}\{G_{\widetilde 3 3}(\omega \ge 0)\} & =&&
-\frac{N_c T\, (\omega^2-\mu\left|\omega\right|)}{8\pi \left[
\sqrt{2\mu\left|\omega\right|}+q^2/(2\pi T) \right]}\,,
\end{align}
and
\begin{equation}
\label{eq:imGX3X3t<}
\mathrm{Im}\{G_{3 \widetilde 3}(\omega  <  0)\} = 0,\qquad
\mathrm{Im}\{G_{\widetilde 3 3}(\omega \ge 0)\} =0.
\end{equation}

\section{Thermal spectral functions}
\label{sec:thermalSpectralFunctions}

We include here a comparision of the sizes of spectral functions
in distinct flavor- and
Lorentz-directions~(see also~(\ref{eq:spectralFunction})
in section~\ref{sec:hydroAdS}).
\begin{figure}[!h]
	\includegraphics[width=.45\linewidth]{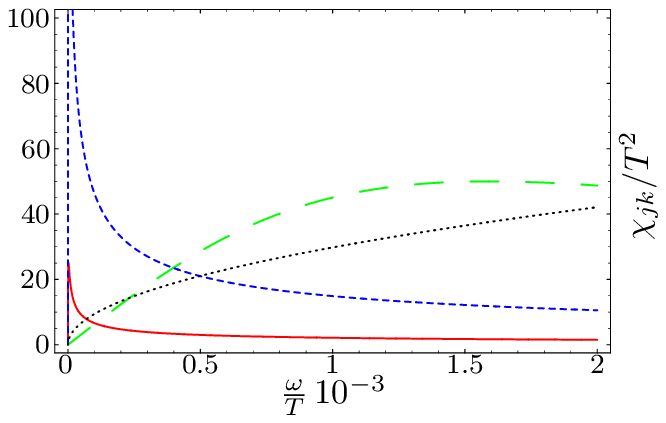}
	\hfill
	\includegraphics[width=.45\linewidth]{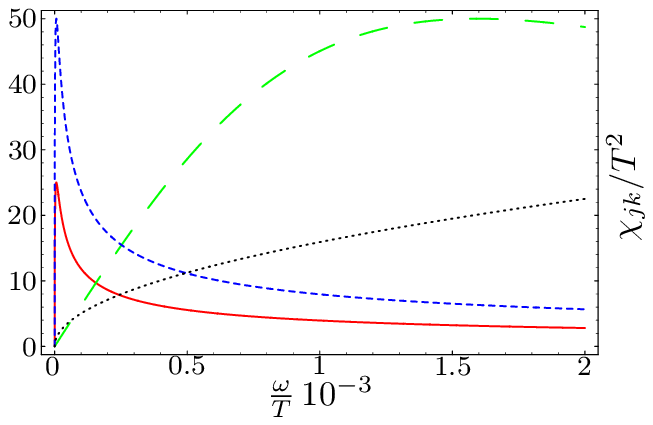}
	\caption{
		Here the thermal spectral functions in distinct Lorentz-
		and flavor-directions are plotted against frequency~$\omega/T$
		in units of temperature. In the left plot the chemical potential
		was chosen to be~$\mu/T=0.7$, in the right one~$\mu/T=0.2$.
		Flavor-directions~$a=1,\,2$ are summed
		and displayed as one curve. The frequency-dependence of
		$00$-~(solid red line) and $03$-Lorentz-directions~(short-dashed blue line)
		is shown. By the dotted line we denote the spectral curve in $11$- or
		$22$-directions. This curve was scaled by a factor 100
		in order to make it visible in these plots. The third flavor-direction is
		only plotted for the spectral function in Lorentz-directions
		$00$~(long-dashed green curve). We do not show the $33$-direction spectral
		function which has a square root dependence and is comparable
		in size with the $11$-direction.
	}
	\label{fig:thermalSpectralFunctions}
\end{figure}

\end{appendix}
\newpage

\providecommand{\href}[2]{#2}\begingroup\raggedright\endgroup

\end{document}